\documentclass[twocolumn,twocolappendix]{aastex701} %linenumbers, trackchanges
\usepackage{amsmath,amssymb}
\usepackage{bm}

\usepackage{algpseudocode}
\newcounter{algorithmN}
\newcommand{\algorithmname}{Algorithm}
\newenvironment{ruledalgorithm}[2][]{%
  \refstepcounter{algorithmN}%
  \par\addvspace{\bigskipamount}%
  \noindent\hrule height 0.8pt\relax
  \vspace{2pt}%
  \noindent\textbf{\algorithmname~\thealgorithmN. #2}%
  \if\relax\detokenize{#1}\relax\else\label{#1}\fi
  \vspace{2pt}\par
  \noindent\hrule height 0.5pt\relax
  \vspace{4pt}%
  \begin{algorithmic}[1]\small
}{%
  \end{algorithmic}%
  \vspace{2pt}%
  \noindent\hrule height 0.8pt\relax
  \par\addvspace{\bigskipamount}%
}
\newcommand{\Given}[1]{\State{\bf given} {#1}}
\newcommand{\RepeatFor}[1]{\Repeat {\bf~for} {#1}}

\usepackage{booktabs}
\usepackage{multirow}
\usepackage{adjustbox}

\usepackage{graphicx}
\usepackage{tikz}

\shorttitle{HyperAIRI}
\shortauthors{C. Tang et al.}
\begin{document}

\title{HyperAIRI: a plug-and-play algorithm for precise hyperspectral image reconstruction in radio interferometry}

\author[orcid=0009-0001-3611-2229,gname=Chao, sname=Tang]{Chao Tang} 
\affiliation{University of Edinburgh, EPCC}
\affiliation{Heriot-Watt University, Institute of Sensors, Signal, and Systems}
\email{c.tang@ed.ac.uk}

\author[orcid=0000-0002-7903-3619,gname=Arwa, sname=Dabbech]{Arwa Dabbech} 
\affiliation{Heriot-Watt University, Institute of Sensors, Signal, and Systems}
\email{a.dabbech@hw.ac.uk}

\author[orcid=0000-0003-0073-682X,gname=Adrian, sname=Jackson]{Adrian Jackson} 
\affiliation{University of Edinburgh, EPCC}
\email{a.jackson@epcc.ed.ac.uk}

\correspondingauthor{Yves Wiaux}
\author[orcid=0000-0002-1658-0121,gname=Yves, sname=Wiaux]{Yves Wiaux} 
\affiliation{Heriot-Watt University, Institute of Sensors, Signal, and Systems}
\email[show]{y.wiaux@hw.ac.uk}

%% Use the \collaboration command to identify collaborations. This command
%% takes an optional argument that is either a number or the word "all"
%% which tells the compiler how many of the authors above the command to
%% show. For example "\collaboration[all]{(DELVE Collaboration)}" wil include
%% all the authors above this command.
%%
%% Mark off the abstract in the ``abstract'' environment. 
\begin{abstract}

The next-generation radio-interferometric (RI) telescopes require imaging algorithms capable of forming high-resolution high-dynamic-range images from large data volumes spanning wide frequency bands. Recently, AIRI, a plug-and-play (PnP) approach taking the forward-backward algorithmic structure (FB), has demonstrated state-of-the-art performance in monochromatic RI imaging by alternating a data-fidelity step with a regularization step via learned denoisers. In this work, we introduce HyperAIRI, its hyperspectral extension, underpinned by learned hyperspectral denoisers enforcing a power-law spectral model. For each spectral channel, the HyperAIRI denoiser takes as input its current image estimate, alongside estimates of its two immediate neighboring channels and the spectral index map, and provides as output its associated denoised image. To ensure convergence of HyperAIRI, the denoisers are trained with a Jacobian regularization enforcing non-expansiveness. To accommodate varying dynamic ranges, we assemble a shelf of pre-trained denoisers, each tailored to a specific dynamic range. At each HyperAIRI iteration, the spectral channels of the target image cube are updated in parallel using dynamic-range-matched denoisers from the pre-trained shelf. The denoisers are also endowed with a spatial image faceting functionality, enabling scalability to varied image sizes. Additionally, we formally introduce Hyper-uSARA, a variant of the optimization-based algorithm HyperSARA, promoting joint sparsity across spectral channels via the $\ell_{2,1}$-norm, also adopting FB. We evaluate HyperAIRI's performance on simulated and real observations. We showcase its superior performance compared to its optimization-based counterpart Hyper-uSARA, CLEAN's hyperspectral variant in WSClean, and the monochromatic imaging algorithms AIRI and uSARA. HyperAIRI's MATLAB implementation is available in the \href{https://basp-group.github.io/BASPLib/}{BASPLib} code library. 

\end{abstract}

%% Keywords should appear after the \end{abstract} command. 
%% The AAS Journals now uses Unified Astronomy Thesaurus (UAT) concepts:
%% https://astrothesaurus.org
%% You will be asked to selected these concepts during the submission process
%% but this old "keyword" functionality is maintained in case authors want
%% to include these concepts in their preprints.
%%
%% You can use the \uat command to link your UAT concepts back its source.
\keywords{\uat{Astronomy image processing}{2306} -- \uat{Computational methods}{1965} -- \uat{Neural networks}{1933} --- \uat{Radio interferometry}{1346}}

%% From the front matter, we move on to the body of the paper.
%% Sections are demarcated by \section and \subsection, respectively.
%% Observe the use of the LaTeX \label
%% command after the \subsection to give a symbolic KEY to the
%% subsection for cross-referencing in a \ref command.
%% You can use LaTeX's \ref and \label commands to keep track of
%% cross-references to sections, equations, tables, and figures.
%% That way, if you change the order of any elements, LaTeX will
%% automatically renumber them.

\section{Introduction}

Radio interferometry is a powerful technique in astronomy, providing high-resolution views of the sky by combining signals from an array of antennas. 
In particular, the upcoming Square Kilometre Array \citep[\href{https://www.skao.int/}{SKA},][]{scaife2020big} is designed to deliver unprecedented sensitivity and resolution, with phase 1 offering more than a tenfold improvement compared to its precursor instruments, \href{https://www.sarao.ac.za/science/meerkat/}{MeerKAT} \citep{jonas2009meerkat} and Australian SKA Pathfinder
\citep[\href{https://www.atnf.csiro.au/facilities/askap-radio-telescope/}{ASKAP,}][]{hotan2021australian}. These telescopes observe across wide frequency bands by collecting Fourier measurements, enabling the study of astrophysical processes at both spatial and spectral scales. 
 
However, the associated inverse problem of hyperspectral image reconstruction is inherently ill-posed due to incomplete Fourier sampling, instrumental noise, and calibration errors. 
Monochromatic algorithms, which reconstruct spectral channels independently, achieve only limited recovery under such conditions. In contrast, joint-channel reconstruction exploits spectral correlations to regularize the ill-posed problem and enables high-precision recovery, particularly in high-dynamic-range (HDR) scenarios.
Nevertheless, the sheer scale of hyperspectral data and the need for precise imaging significantly increase the computational demands, calling for efficient and scalable algorithms with joint-channel reconstruction capability.
 
Most radio sources emit synchrotron radiation that can be described by a power-law spectral model \citep{thompson2017interferometry}. Building on the celebrated CLEAN algorithm \citep{hogbom1974aperture}, its first hyperspectral extension, Multi-Frequency (MF) CLEAN, models spectral variations via polynomial functions, and solves for the corresponding coefficient images during reconstruction \citep{sault1994multi}. Further improvements are enabled by combining MF with Multi-Scale (MS) CLEAN \citep{cornwell2008multiscale}, enhancing the recovery of extended emission \citep{rau2011multi, offringa2017optimized}.
However, low-order polynomials are often inadequate to capture complex spectral behavior \citep{scaife2012broad}.
To address this, \citet{ceccotti2023novel} have injected a forced-spectrum fitting approach within CLEAN, enforcing a prior spectral knowledge directly during reconstruction.
While computationally efficient, CLEAN-based algorithms still fall short in achieving precise imaging, \textit{i.e.} high-resolution and high-sensitivity imaging. These limitations stem from their simplistic prior model: spatial resolution remains constrained by the instrument's nominal resolution, and dynamic range is degraded by the addition of residuals in the final reconstruction.

Bayesian inference methods, such as the variational Bayes approach underpinning the RESOLVE algorithm \citep{junklewitz2016resolve}, have demonstrated superior precision and robustness over CLEAN-based techniques \citep{arras2021comparison, roth2023bayesian}. Assuming a first-order power-law spectral model,  \citet{junklewitz2015new} proposed the joint estimation of a reference-frequency image and a spectral index map. Although these methods naturally provide uncertainty quantification functionality, their scalability and precision on large-scale observations remain to be demonstrated.

Methods rooted in optimization theory have shown promise in astronomical imaging, offering precision imaging. Optimization provides a robust and versatile framework to address ill-posed inverse problems, typically formulated as minimization tasks solved via convergent iterative algorithmic structures.
In this context, the reconstructed image is obtained as the minimizer of an objective function composed of a data fidelity function and a regularization function encoding sophisticated image models.
Optimization offers a multitude of distributed algorithmic structures for parallelized reconstruction at scale. Among these methods, the SARA family of algorithms leverages sparsity-based priors, demonstrating superior imaging quality compared to CLEAN across diverse imaging scenarios \citep{carrillo2012sparsity,onose2016scalable,onose2017accelerated,repetti2017non,birdi2018sparse,pratley2018robust,dabbech2018cygnus,terris2023image,wilber2023scalableI}. Specifically, HyperSARA, promotes joint-channel average sparsity of the hyperspectral image cube in a redundant wavelet dictionary, and its low-rankness \citep{abdulaziz2016low, abdulaziz2019wideband}.
HyperSARA takes the primal-dual forward-backward algorithmic structure \citep[PD-FB,][]{condat2013primal}, which enables full splitting of the functions involved, and parallel updates of the dual variables. 
To further improve scalability to large image sizes, \citet{thouvenin2023parallelI} proposed its faceted variant operating on 3D spatial-spectral image facets. The algorithm has been validated on real observational data, demonstrating state-of-the-art reconstruction \citep{thouvenin2023parallelII}.
Yet, HyperSARA requires heavy cross-channel communications and could introduce significant overhead for extreme-scale problems.
In this work, we introduce another simpler variant of HyperSARA, dubbed Hyper-uSARA, which adopts an unconstrained formulation of the minimization task, thereby eliminating the need for exact noise distribution knowledge.
It also takes a simpler algorithmic structure, the FB \citep{combettes2011proximal}, alternating data fidelity and regularization steps, while still allowing for full parallelization within each step.

Recently, deep learning has gained increasing attention in RI imaging for its ability to produce reconstructions with rapid inference. Existing approaches focus on monochromatic imaging. These include traditional, end-to-end neural networks, either reconstructing images from back-projected data \citep{connor2022deep, gheller2022convolutional} or post-processing suboptimal outputs from other algorithms such as CLEAN \citep{terris2019deep}. However, these methods often lack precision, robustness, and interpretability \citep{goodfellow2014explaining, nguyen2015deep, pang2018towards}. 
Conditioned denoising diffusion probabilistic models have also been explored for RI imaging \citep{wang2023conditional, drozdova2024radio}. They face similar issues and remain in early development. 

Advanced architectures such as unrolling Deep Neural Networks (DNN) promote data consistency and possibly algorithmic stability by unfolding a fixed number of iterations of an optimization algorithm as key modules within the network \citep{monga2021algorithm}. These networks are trained end-to-end while preserving an explicit link to optimization theory.
However, existing unrolling schemes remain limited by their fixed iteration count, and lack formal convergence guarantees.
Furthermore, embedding computationally-intensive measurement operators directly into the network can make training prohibitively expensive. 

In contrast, the Plug-and-Play (PnP) framework bridges optimization theory and deep learning \citep{venkatakrishnan2013plug, romano2017little, reehorst2018regularization} offering a more flexible and robust alternative. In this approach, proximal operators enforcing handcrafted regularization within an optimization algorithm are replaced by learned DNN denoisers. These denoisers can be agnostic to the measurement process, making them adaptable to diverse configurations in RI imaging. \added{Among these are the AIRI family of PnP algorithms \citep{terris2023image, terris2025airi} which are underpinned by learned denoisers trained via a tailored approach, endowing them with convergence guarantees \citep{pesquet2021learning}}. AIRI has been validated on both simulated and real wide-field observational data \citep{dabbech2022first, wilber2023scalableII}. Thanks to their convolutional architecture and compact receptive fields, AIRI denoisers are typically trained on small image patches, and seamlessly applied to images of arbitrary size during inference. Additionally, they can be easily applied in parallel to image facets for scalability. These properties make AIRI well suited to large-scale precision imaging.

Inspired by unrolled networks and PnP algorithms, 
\citet{aghabiglou2024r2d2} have recently proposed R2D2, a novel imaging paradigm for radio interferometry. R2D2 is underpinned by a series of iteration-specific DNNs, each taking as input the current image estimate and associated data residual. The final reconstruction is formed by summing the DNNs' output. In contrast to unrolled networks, R2D2 decouples data fidelity from the network architecture thereby avoiding the associated computational challenges. Conceptually, it can be interpreted as a learned version of CLEAN \citep{dabbech2024cleaning}, substituting CLEAN's minor cycles with iteration-specific DNNs. The novel paradigm has demonstrated superior imaging precision to the state of the art in field, namely AIRI and uSARA, both in simulation and on real data, only at a fraction of their computational cost. However, early incarnations of R2D2 have been so far trained in a telescope-specific setting, and for monochromatic intensity imaging at small scale. Extending it to large scale wideband imaging poses additional challenges that remain to be addressed.

Building on the precision and scalability of AIRI, we propose HyperAIRI, a novel PnP algorithm tailored for high-precision hyperspectral RI imaging. HyperAIRI applies learned denoisers to each spectral channel in parallel with side information from neighboring channels. The reconstruction approach further enhances spectral coherence by incorporating a physical power-law model, with the spectral index map either estimated on the fly during reconstruction or provided as an input. This design ensures local communication at each iteration and requires global communication only during spectral index updates, making HyperAIRI highly scalable and efficient in distributed high-performance computing (HPC) environments. Similarly to AIRI, HyperAIRI takes FB, with convergence guaranteed by the non-expansiveness of the denoisers, enforced via a Jacobian regularization term in the training loss. We provide a shelf of pre-trained denoisers covering a wide dynamic-range, eliminating the need for problem-specific training. We demonstrate the superior performance of HyperAIRI, compared with state-of-the-art methods in the field, including the optimization-based Hyper-uSARA and joint-channel CLEAN, on simulated the Very Large Array \citep[\hyperlink{https://public.nrao.edu/telescopes/vla/}{VLA},][]{perley2011expanded} data and real ASKAP observations.

The remainder of this paper is organized as follows. In Section~\ref{sec:theory}, we formalize the hyperspectral RI imaging problem.  Section~\ref{sec:hyperusara} reviews the hyperspectral SARA prior and formulates the Hyper-uSARA algorithm. Section~\ref{sec:hyperairi} presents the HyperAIRI algorithm, including the network architecture, training methodology, and overall algorithm structure. Section~\ref{sec:sim_exp} reports experimental results on simulated datasets to assess HyperAIRI's performance under controlled conditions. Section~\ref{sec:askap} demonstrates its application to real ASKAP observations. Finally, Section~\ref{sec:conclusion} summarizes the contributions and discusses potential directions for future work.

%%%%%%%%%%%%%%%%%%%%%%%%%%%%%%%%%%%%%%%%%%%%%%
%%%%%%%%%%%%%%%%%%%%%%%%%%%%%%%%%%%%%%%%%%%%%%
%%%%%%%%%%%%%%%%%%%%%%%%%%%%%%%%%%%%%%%%%%%%%%

\section{Hyperspectral RI imaging} \label{sec:theory}

In this section, we present the power-law spectral model of the radio emission, examine the specificity of the hyperspectral Fourier sampling, and formulate the corresponding measurement model that defines the RI inverse problem. 

\subsection{Hyperspectral radio emission model}

In radio astronomy, continuum synchrotron emission typically varies smoothly with frequency, with flux density scaling according to a power-law \citep{thompson2017interferometry}. This makes the power-law spectral model a practical choice for parameterizing the hyperspectral RI image cube. To account for sudden changes in the spectrum
while still preserving overall spectral smoothness across a wide frequency range, a second-order term is introduced in the exponent \citep{rau2011multi}.
Consider a reference spectral frequency $\nu_0$ and the corresponding channel image $\overline{\bm{x}}_0 \in \mathbb{R}^N$. At a given frequency $\nu_l$, the radio image $\overline{\bm{x}}_l \in \mathbb{R}^N$ can be modeled as
\begin{equation}
    \overline{\bm{x}}_l = \overline{\bm{x}}_0 \left ( {\nu_l}/{\nu_0} \right ) ^{ -\bm{\alpha} + \bm{\beta} \log \left ( {\nu_l}/{\nu_0} \right ) }
    \label{eq:power_law}
\end{equation}%
where $\bm{\alpha} \in \mathbb{R}^N$ is the spectral index map and $\bm{\beta} \in \mathbb{R}^N$ is the spectral curvature map.

This model is widely used for characterizing various radio sources, including galaxies, which yield different spectral indexes due to the different physical radiation mechanisms \citep{de2018radio}. Furthermore, it can serve as a physics-informed prior in the reconstruction of the hyperspectral image cube, ensuring meaningful correlations across different spectral channels
\citep{junklewitz2015new, ceccotti2023novel}.

\subsection{Hyperspectral Fourier sampling in RI}

\begin{figure}
    \centering
    \begin{tabular}{c@{\hspace{0.\tabcolsep}}c@{\hspace{0.\tabcolsep}}c}
        \includegraphics[width=0.32\linewidth]{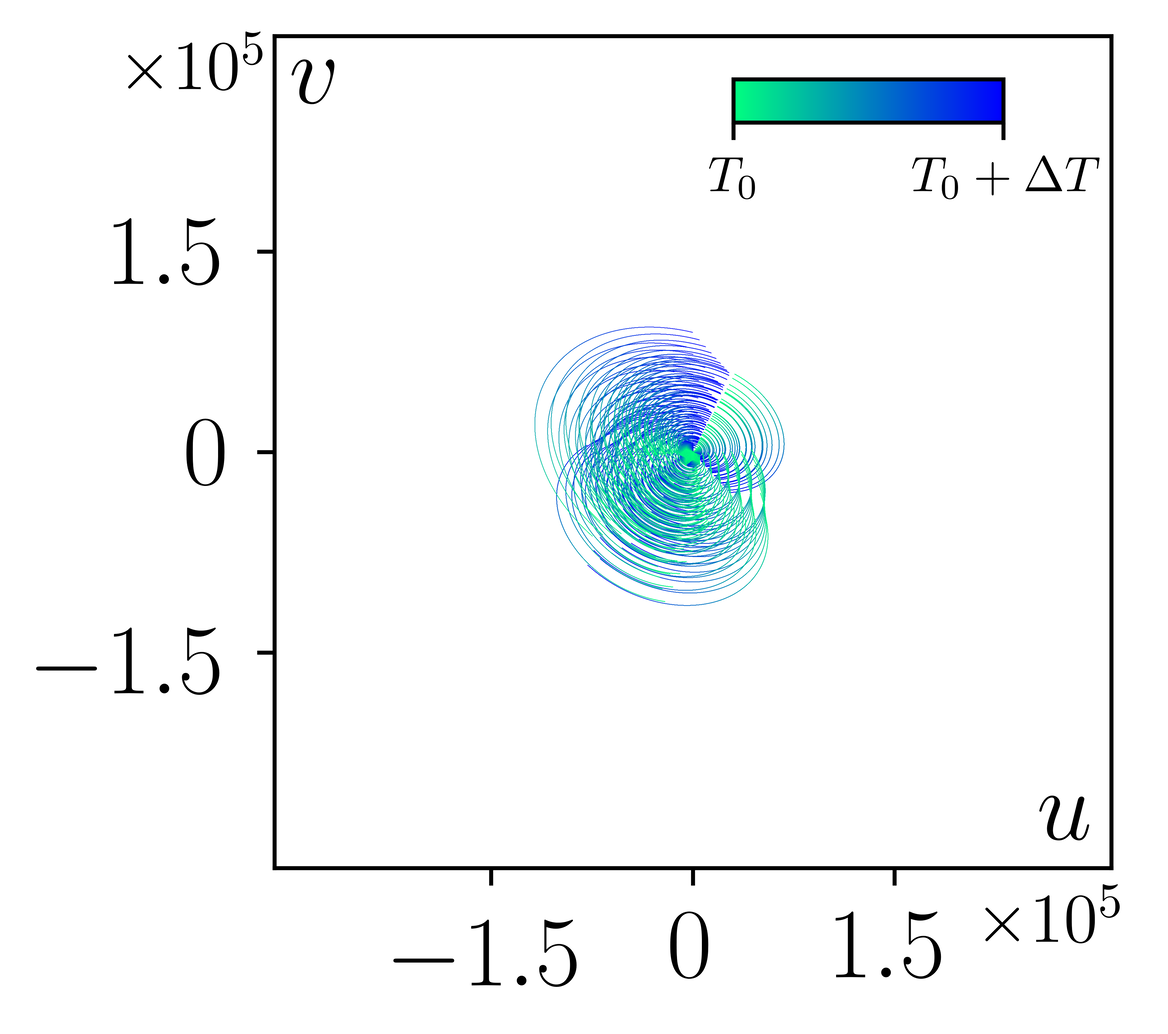}
        &
        \includegraphics[width=0.32\linewidth]{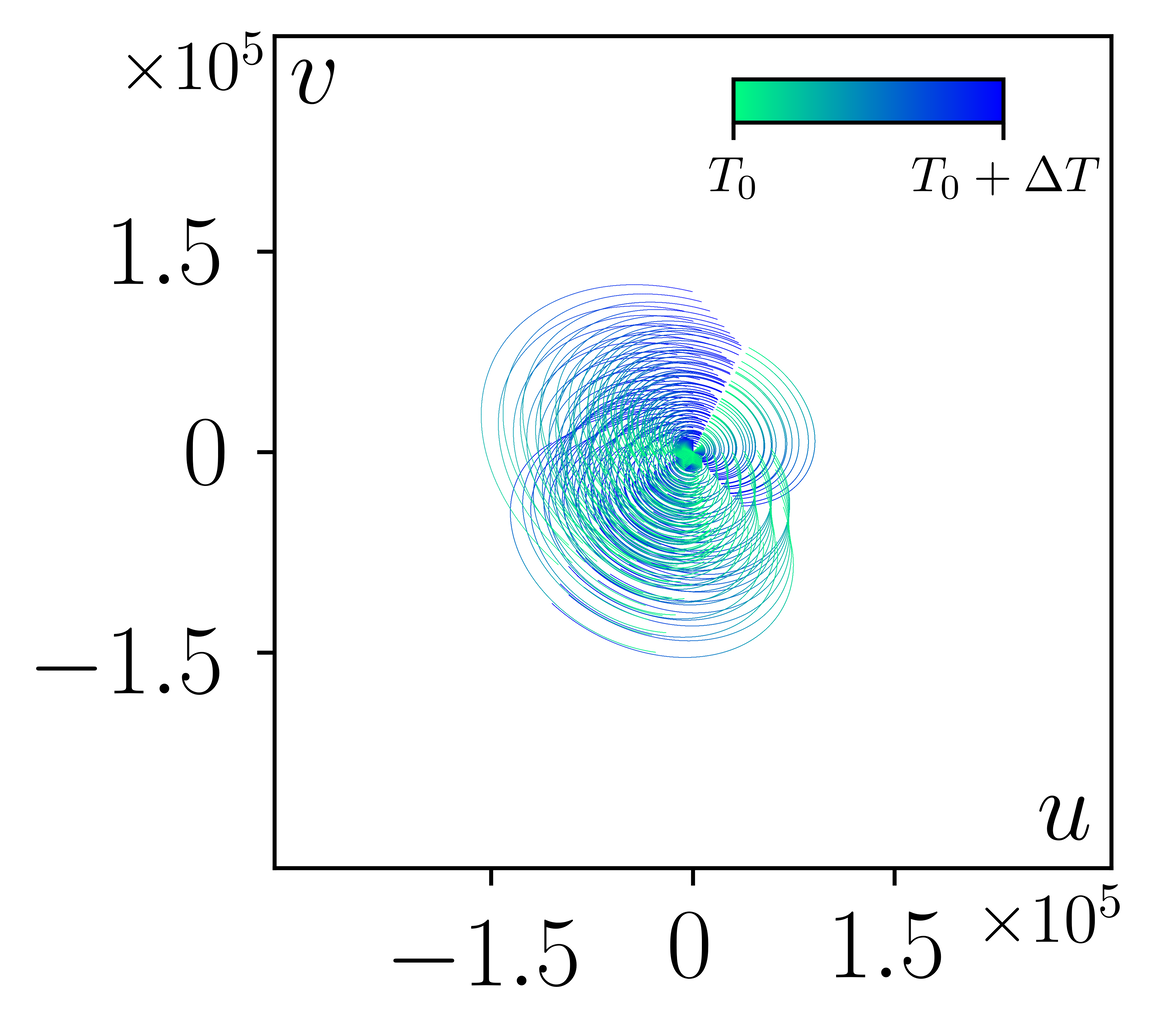}
        &  
        \includegraphics[width=0.32\linewidth]{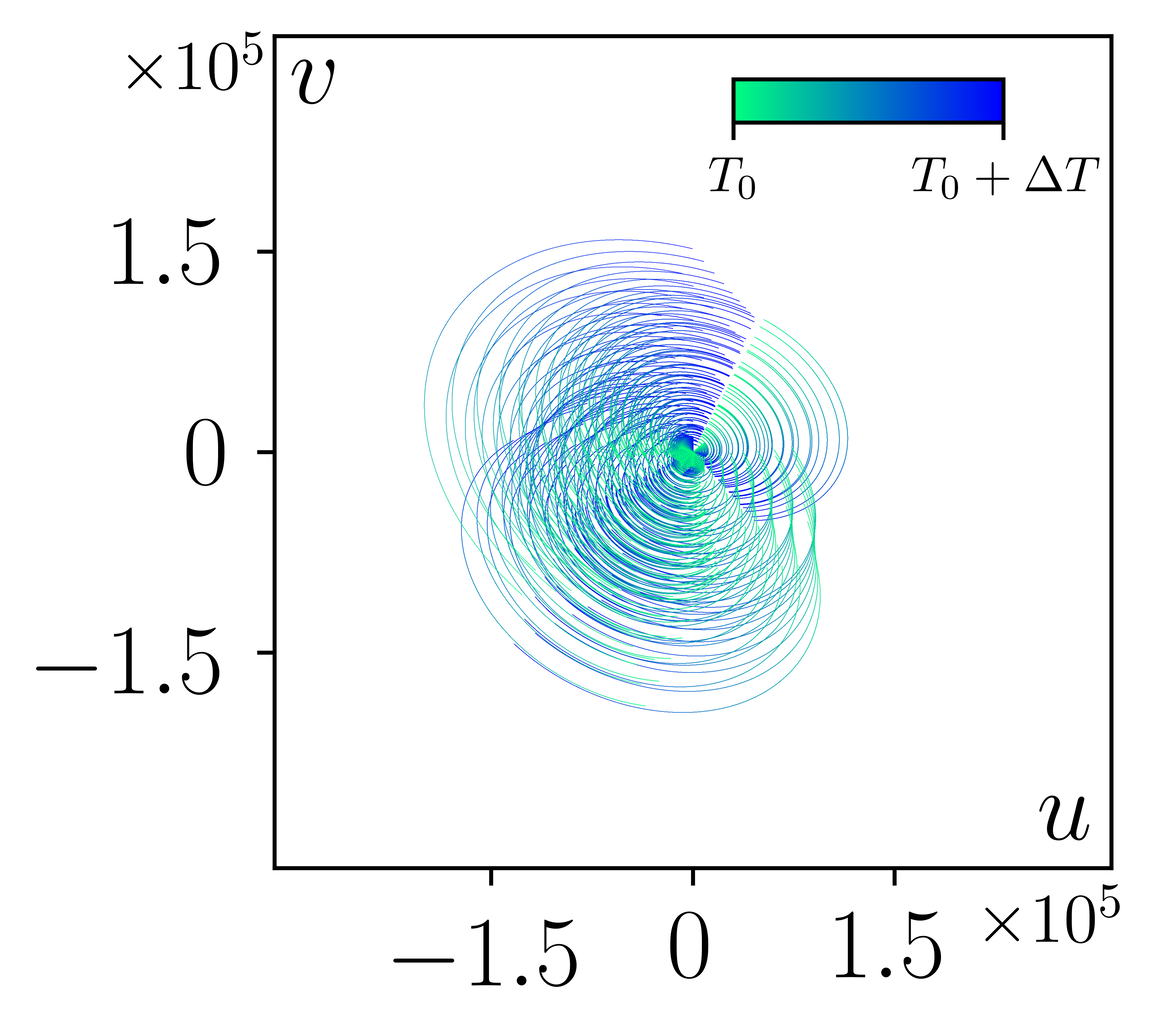}
        \\
       (a) channel 1 & (b) channel 18 & (c) channel 36
    \end{tabular}
    \caption{A set of simulated VLA hyperspectral sampling patterns of different channels. The three panels show the normalized $uv$ sampling patterns for the first, middle and last channels of an observation with frequencies 1.00, 1.34 and 1.70 GHz, respectively. The coordinates are in the unit of wavelength.}
    \label{fig:sim_uv}
\end{figure}

When probing the radio sky, RI measurements are acquired using arrays of antennas operating at multiple spectral channels.
At a given observation frequency, each pair of antennas captures a noisy spatial Fourier intensity measurement, also known as a visibility. 
The corresponding Fourier mode is derived from the projection of the associated baseline onto the plane orthogonal to the line of sight, denoted by $\overline{\bm{b}}$ in the unit of meters
\citep{thompson2017interferometry}.
The collection of visibilities from all baselines, accumulated over time, forms the so-called $uv$-coverage.
For a given channel $l \in \{1, \dots, L\}$, the probed Fourier mode $\bm{b}_l$ is described in the unit of the observation wavelength such that
\begin{equation}
    \bm{b}_l = \frac{\nu_l}{c} \overline{\bm{b}},
\end{equation}%
where $\nu_l$ is the observation frequency, and $c$ is the speed of light. 
Consequently, for a fixed set of baselines, the $uv$-coverage becomes more concentrated towards the center of the Fourier plane, at the lower end of the observation frequency bandwidth. The same sampling pattern dilates radially away from the center of the Fourier plane, at the higher end of the observation frequency bandwidth.
Figure~\ref{fig:sim_uv} shows this spectral-frequency-dependent dilation of the $uv$-sampling pattern across three representative channels using a simulated VLA $uv$-coverage. By design, observations at lower-frequency channels provides  lower spatial resolution but higher sensitivity, and conversely, higher-frequency channels offer higher resolution but lower sensitivity due to the scaling of the $uv$-coverage.

\subsection{RI hyperspectral inverse problem} \label{sec:HRI_problem}
Focusing on intensity imaging and assuming a narrow field of view without atmospheric or instrumental perturbations,  at a given spectral channel $l$,
the process of measuring visibilities $\bm{y}_l \in \mathbb{C}^{M}$ from the true sky intensity image ${\overline{\bm{x}}}_l \in \mathbb{R}^{N}_+$ can be modeled by the following discrete linear system:
\begin{equation}
    \bm{y}_l = \bm{\Phi}_l {\overline{\bm{x}}}_l + \bm{n}_l, \; \mathrm{with} \, \bm{\Phi}_l = \mathbf{G}_l \mathbf{F P},
    \label{eq:model_single}
\end{equation}%
where $\bm{\Phi}_l \in \mathbb{C}^{M \times N}$ represents the measurement operator at channel $l$, and $\bm{n}_l \in \mathbb{C}^M$ denotes measurement noise, modeled as a complex white Gaussian noise with mean zero and a standard deviation of $\tau_l$. The measurement operator $\bm{\Phi}_l$ models incomplete non-uniform Fourier sampling where ${\mathbf{G}_l} \in \mathbb{C}^{M \times N'}$ is a sparse non-uniform Fourier interpolation matrix, with each row corresponding to a compact convolutional interpolation kernel centered at the corresponding Fourier mode, \added{${\mathbf{F}} \in \mathbb{C}^{N' \times N'}$} denotes the 2D Fourier transform, and ${\mathbf{P}} \in \mathbb{R}^{N' \times N}$ is a zero-padding operator, also encompassing the correction for the convolutional interpolation kernels \citep{fessler2003nonuniform}.

In practice, the noise $\bm{n}_l$ may be non-white, in which case a diagonal noise-whitening matrix $\mathbf{\Theta}_l \in \mathbb{R}^M_+$ is incorporated into $\mathbf{\Phi}_l$ and applied to the measurements $\bm{y}_l$, such that
$\bm{\Phi}_l = \mathbf{\Theta}_l \mathbf{G}_l \mathbf{F P}$. Each diagonal element of $\mathbf{\Theta}_l$ is the inverse of the noise standard deviation of the corresponding visibility. With this adjustment known as natural weighting, the linear measurement model in \eqref{eq:model_single} remains valid for the whitened visibilities $\bm{y}_l$, whereby $\tau_l = 1$. 
On a further note, the RI measurement process can be affected by the so-called  $w$-effect, a phase modulation arising from the non-coplanarity of the antenna array which is non-negligible in wide-field imaging. In this context, the operator $\bm{\Phi}_l$ can account for this effect via additional row-wise convolutions \citep{dabbech2017w}.

In what follows, the hyperspectral visibilities, image cube, and measurement noise are arranged in a matrix form as $\mathbf{Y} = (\bm{y}_1,...,\bm{y}_L) \in \mathbb{R}^{M \times L}$, $\mathbf{X} = (\bm{x}_1,...,\bm{x}_L) \in \mathbb{R}^{N \times L}$, and $\mathbf{N} = (\bm{n}_1,...,\bm{n}_L) \in \mathbb{R}^{M \times L}$ respectively. The hyperspectral measurement operator is defined such that ${\Phi} ( {\mathbf{X}} ) = ([\bm{\Phi}_l {\bm{x}}_l]_{1 \leq l \leq L})$. Using this notation, the hyperspectral RI measurement model becomes: 
\begin{equation} 
    \mathbf{Y} = {\Phi} ( {\mathbf{X}} ) + \mathbf{N}. \label{eq:model_wideband} 
\end{equation}%
Hyperspectral RI imaging aims to jointly reconstruct
the spatial and spectral information of the radio emission encapsulated in the hyperspectral cube $\mathbf{X}$ from the incomplete and noisy measurements $\mathbf{Y}$. 
It is evident from \eqref{eq:model_single} that the hyperspectral RI measurement process is separable across channels, allowing monochromatic algorithms to reconstruct each spectral channel independently. However, joint spectral imaging methods can yield higher-quality reconstructions by leveraging the higher sensitivity of lower-frequency channels and the improved resolution of higher-frequency channels.

%%%%%%%%%%%%%%%%%%%%%%%%%%%%%%%%%%%%%%%%%%%%%%
%%%%%%%%%%%%%%%%%%%%%%%%%%%%%%%%%%%%%%%%%%%%%%
%%%%%%%%%%%%%%%%%%%%%%%%%%%%%%%%%%%%%%%%%%%%%%

\section{Optimization-based hyperspectral imaging} \label{sec:hyperusara}
 
Building on HyperSARA, we introduce its unconstrained variant, Hyper-uSARA, tailored for hyperspectral imaging with unknown noise statistics and calibration errors. This variant is particularly well suited to large-scale real observational data, where flexibility and efficiency are critical.
We begin by revisiting the optimization theory underpinning our method, focusing on the FB for solving unconstrained minimization problems. We then describe the hyperspectral SARA prior, which promotes joint-channel average sparsity, and present the full algorithmic structure of Hyper-uSARA.

\subsection{Optimization theory}

The linear measurement model taking the form \eqref{eq:model_wideband} is widely used across various imaging tasks, such as denoising, super-resolution, inpainting, image deblurring, and magnetic resonance imaging. While the measurement operator $\Phi(\cdot)$ varies across image modalities, the underlying inverse problem is often ill-posed due to incomplete and noisy measurements. 
Optimization methods are thus appealing for their versatility in regularizing inverse problems by injecting sophisticated image models.

Within the convex optimization framework, the inverse problem is addressed by solving a minimization problem with objective function
\begin{equation}
    \underset{\mathbf{X} \in \mathbb{R}^{N \times L}}{\min} \; f(\mathbf{X}) + \lambda r(\mathbf{X}) ,
    \label{eq:obj}
\end{equation}%
where $f : \mathbb{R}^{N \times L} \mapsto \mathbb{R} \in \Gamma_0(\mathbb{R}^{N \times L})$\footnote{$\Gamma_0(\mathbb{R}^{N \times L})$ denotes the set of convex, proper, and lower semi-continuous functions from $\mathbb{R}^{N \times L}$ to $(-\infty, \infty)$.} is the data-fidelity term reflecting the linear measurement model, $r : \mathbb{R}^{N \times L} \mapsto \mathbb{R} \in \Gamma_0(\mathbb{R}^{N \times L})$ is the regularization term enforcing the image model. 
Given the Gaussian nature of the measurement noise, a squared-error fidelity term is one common choice for $f$, such that $f(\mathbf{X}) = \frac{1}{2} \| \Phi ( \mathbf{X} ) - \mathbf{Y} \|^2_{\mathrm{F}}$, where $ \| \cdot \|_F $ denotes Frobenius norm. 
The parameter $\lambda > 0$ is a regularization parameter balancing the two terms. 
Inserting $f$ into \eqref{eq:obj} yields an unconstrained minimization task.

When at least one of the two terms is differentiable--in this case the data fidelity term $f$--problems of the form \eqref{eq:obj} can be solved  with FB.
The iteration structure consists of a gradient descent step (forward) for the differentiable data-fidelity term, followed by and denoising step (backward) via the proximity operator of the non-differentiable regularization term. At any iteration $i$, the image cube update reads
\begin{equation}
    \mathbf{X}^{i+1} = \mathrm{prox}_{\gamma \lambda r} \big( \mathbf{X}^i - \gamma \nabla f \big(\mathbf{X}^i \big) \big).
    \label{eq:fb}
\end{equation}%
The gradient of $f(\cdot)$ reads  $\nabla f \left (\mathbf{X} \right ) = \mathrm{Re} \{ \Phi^\dagger \left ( \Phi \left (\mathbf{X} \right )  \right ) \} - \mathrm{Re} \{ \Phi^\dagger (\mathbf{Y}) \}$, where $\Phi^\dagger (\cdot)$ is the adjoint of hyperspectral measurement operator and $ \mathbf{X}^{\mathrm{dirty}} = \mathrm{Re} \{ \Phi^\dagger (\mathbf{Y}) \}$ is the back-projected data, commonly referred to as the dirty image cube. \added{We further define the residual dirty image cube as $\mathbf{X}^{\mathrm{res}} = \mathbf{X}^{\mathrm{dirty}} - \mathrm{Re}\{ \Phi^\dagger(\Phi(\mathbf{X})) \}$, from which it follows that $\nabla f(\mathbf{X}) = -\,\mathbf{X}^{\mathrm{res}}$. Also, the channel-wise dirty image and residual dirty image are expressed as $\bm{x}_l^{\mathrm{dirty}} = \mathrm{Re} \{ \bm{\Phi}^\dagger \bm{y}_l \}$ and $\bm{x}_l^{\mathrm{res}} = \bm{x}_l^{\mathrm{dirty}} - \mathrm{Re} \{ \bm{\Phi}^\dagger \bm{\Phi} \bm{x}_l\}$, respectively.}
The parameter $\gamma$ is the step size, upper-bounded by $ 2/\| \Phi^{\dagger}(\Phi(\cdot)) \|_\mathrm{S}$ to ensure the algorithm's convergence\footnote{In practice, $\gamma$ is typically fixed to $1.98/\| \Phi^{\dagger}(\Phi(\cdot)) \|_\mathrm{S}$.}, with $\| \cdot\|_\mathrm{S}$ denoting the spectral norm \citep{bauschke2017Convex}. 

By definition, the proximity operator of a regularization function $r$ is given by
\begin{equation}
    \mathrm{prox}_r (\mathbf{Z}) = \underset{\mathbf{U} \in \mathbb{R}^{N \times L} }{\mathrm{argmin}}  \; \frac{1}{2} \| \mathbf{U} - \mathbf{Z} \|^2_{\mathrm{F}} + r(\mathbf{U}).
    \label{eq:prox}
\end{equation}%
When $r$ is separable across channels, \textit{i.e.} taking the form $r(\mathbf{X}) = \sum_{l=1}^{L} r_l(\bm{x}_l)$, naturally the problem \eqref{eq:obj} decomposes into $L$ independent monochromatic imaging problems. To exploit spectral correlations and improve reconstruction quality, joint regularization terms are preferred for hyperspectral RI imaging.

\subsection{Hyperspectral SARA prior and Hyper-uSARA}

The choice of the prior encoded via the regularization function $r$ in \eqref{eq:obj} greatly affects imaging precision.
In monochromatic RI imaging, the sparsity-based SARA prior promotes average sparsity of the sought image across a collection of 8 orthonormal wavelet bases and the Dirac basis, denoted as $\bm{\Psi} \in \mathbb{R}^{N \times bN}$ with $b=9$.
Its extension to hyperspectral imaging, the HyperSARA prior, is underpinned by the assumption that the image cube is the combination of a few radio sources, each having a distinct spectral behavior \citep{abdulaziz2019wideband}. As such, the HyperSARA prior promotes the low-rankness of the image cube, whose rank is upper-bounded by the number of radio sources, and its joint-sparsity over the spectral channels in the SARA dictionary. The former is enforced via a nuclear norm which penalizes the $\ell_1$-norm of the vector of the singular values of $\mathbf X$. The latter is enforced via an $\ell_{2,1}$-norm, defined as the $\ell_1$-norm of the vector whose elements are the $\ell_2$-norms of the rows of $\bm{\Psi}^\dagger\mathbf{X}$. Focusing on intensity imaging, the prior also includes non-negativity of the target image cube. The HyperSARA algorithm takes a constraint minimization task, involving bounds on the hyperspectral noise cube, and is solved using the PD-FB.     

Underpinned by a constrained data fidelity term, HyperSARA requires accurate knowledge of the noise statistics, which can be challenging in practice, with calibration errors often dominating the theoretical noise. To address this, we formally propose the unconstrained formulation of the HyperSARA minimization task. Furthermore, we discard the low-rankness term, though accepting a slight loss in precision, since it involves singular value decomposition of the image cube estimate at every iteration, a computationally expensive operation for large-scale imaging even when adopting 3D faceting as in Faceted HyperSARA \citep{thouvenin2023parallelI, thouvenin2023parallelII}. 
This simplification also eliminates the need to balance the average joint-sparsity and low-rankness terms, while reducing the number of regularization parameters.
The resulting unconstrained variant of HyperSARA underpinned by an average joint-sparsity prior is dubbed Hyper-uSARA.
The regularization function reads 
\begin{equation}
     \begin{aligned}
          r(\mathbf{X}) = & \sum^{bN}_{n=1}{ 
          \epsilon
          \log \left ( 
          \epsilon^{-1}
          \| ( (\bm{\Psi}^\dagger \mathbf{X})^\top )_n \|_2 + 1 \right ) }
          \\ & \qquad + \iota_{\mathbb{R}_+^{N \times L}} (\mathbf{X}),
     \end{aligned}
    \label{eq:sara_log_prior}
\end{equation}%
where $\epsilon$ is a hyperparameter, $(\cdot)_n$ represent the $n$-th column of the argument matrix, $\| \cdot \|_2$ denotes the $\ell_2$-norm, and $\iota_{\mathbb{R}_+^{N \times L}} (\cdot)$ is an indicator function\footnote{The indicator function $\iota_{\mathbb{R}_+^{N \times L}} (\mathbf{X})=0$ if all elements in $\mathbf{X}$ are positive, otherwise $\iota_{\mathbb{R}_+^{N \times L}} (\mathbf{X})=\infty$.} imposing positivity constraint on each pixel.
To address the non-convex regularization function $r$, the minimization problem of the form \eqref{eq:obj} is addressed via a re-weighting scheme that iteratively solves a sequence of convex minimization tasks.
More precisely, at each re-weighting iteration a minimization task involving a surrogate convex regularization function $\widetilde{r}$ instead of $r$ \citep{repetti2021variable}, such that
\begin{equation}
    \widetilde{r}(\mathbf{X}) = \| \bm{\Psi}^\dagger\mathbf{X} \|_{2,1,\bm{w}} + \iota_{\mathbb{R}_+^{N \times L}} (\mathbf{X}).
    \label{eq:sara_prior}
\end{equation}%
The weighted average joint-sparsity regularizer is given by
\begin{equation}
    \| \bm{\Psi}^\dagger \mathbf{X} \|_{2,1,\bm{w}} 
    = \left \|\mathbf{W}
    \sqrt{\sum_{l=1}^{L} (\bm{\Psi}^\dagger \bm{x}_l )^2 }  \right \|_1,
\end{equation}%
where $\mathbf{W} \in \mathbb{R}^{bN \times bN}$ is a diagonal matrix, whose diagonal elements $w_n$ form the vector $\bm{w} \in \mathbb{R}^{bN}$. The weights $\bm{w}$ are updated at each re-weighting cycle from the solution of the preceding minimization task as
\begin{equation}
    \bm{w} = \epsilon\left/\left(\epsilon+  \sqrt{\sum_{l=1}^{L} \left (\bm{\Psi}^\dagger \bm{x}_l \right )^2 } \right)\right..
    \label{eq:reweight_coeff}
\end{equation}%

At each re-weighting iteration $j$, the objective function is updated using the current $\bm{w}_j$, and is solved via the FB \eqref{eq:fb}. 
The Hyper-uSARA algorithm is summarized in Algorithm~\ref{algo:hyperusara}. 
The proximity operator of the composite regularization function \eqref{eq:sara_prior} is computed via a sub-iterative dual-FB \citep{combettes2011proximal} illustrated in Algorithm~\ref{algo:prox_hyperusara}. 
At a given $\mathbf{Z} \in \mathbb{R}^{N \times L}$, the respective proximity operators for the positivity constraint and the weighted $\ell_{2,1}$-norm sparsity term have closed-form solutions such that
\begin{equation}
    \mathrm{prox}_{\iota_{\mathbb{R}_+^{N \times L}} (\cdot)}(\mathbf{Z}) = \mathrm{max}\{ \mathbf{Z}, \mathbf{0} \},
\end{equation}%
and
\begin{equation}
     \begin{aligned}
          \mathrm{prox}_{\eta \| \cdot \|_{1,2,\bm{w}}}(\mathbf{Z}) &= \Biggl( \Biggl[ \mathrm{max}\!\left\{ \| ( \mathbf{Z}^{\top} )_n\|_2 - \eta w_n, 0 \right\}
          \\[-2pt]
          &\qquad\qquad \frac{ ( \mathbf{Z}^{\top} )_n}{\| (\mathbf{Z}^{\top} )_n\|_2} \Biggr]_{1 \leq n \leq N} \Biggr)^{\!\top}.
     \end{aligned}
\end{equation}%
As in Faceted-HyperSARA, the average joint-sparsity regularization is applied in parallel to each 3D image facet using the same splitting strategy.
Formally, the FB gradient step only requires the dirty image and the application of $\Phi^\dagger(\Phi(\cdot))$, keeping memory and computational cost tied to the image size \citep{vijay2017fourier}.
Hence, Hyper-uSARA can achieve better efficiency for large-scale imaging tasks under tight computational constraints.

Concerning the choice of the regularization parameters involved, both $\lambda$ and $\epsilon$ are inferred from the image-domain noise level.
\citet{terris2023image} proposed estimating the image-domain noise level at each frequency channel as
\begin{equation}
    \sigma_{\mathrm{heu},l} = \left . \tau_l  \middle / \sqrt{2\| {\bm{\Phi}}_l \|^2_{\rm S}} \right ..
    \label{eq:heuristic}
\end{equation}%
Under this consideration, and following \citet{thouvenin2023parallelI}, the noise floor level $\epsilon$ in the $\ell_2$-norm of the wavelet coefficients is then approximated such that
\begin{equation}
    \epsilon^2 = \left . {\sum^L_{l=1} \sigma_{_{\mathrm{heu}, l}}^2 / {b}} \right .,
\end{equation}%
and the regularization parameter is chosen as $\lambda = \epsilon / \gamma$ \citep{thouvenin2023parallelI}.

\begin{ruledalgorithm}[algo:hyperusara]{Hyper-uSARA}
    \State{\textcolor{green!70!black}{// Initialization}}
    \Given{$\mathbf{X}^0$, $\mathbf{X}^{\mathrm{dirty}}$, $\Phi(\cdot)$, $\Phi^\dagger(\cdot)$ }
    \State{\textcolor{green!70!black}{// Step size, regularization parameter, noise floor level}}
    \Given{$\gamma$, $\lambda$, $\epsilon$}
    \State{\textcolor{green!70!black}{// Tolerance for re-weighting and FB}}
    \Given{$\xi_1$, $\xi_2$}
    \State{\textcolor{green!70!black}{// Maximum iterations for re-weighting and FB}}
    \Given{$J$, $I$}
    \State{\textcolor{green!70!black}{// Re-weighting loop}}
    \RepeatFor{$j=1,2,\ldots,J$}
        \State {$\bm{w}^{j} = \epsilon\left/\left(\epsilon+  \sqrt{\sum_{l=1}^{L} \left (\bm{\Psi}^\dagger \bm{x}^{j-1}_l \right )^2 } \right)\right. $}
            \State{\textcolor{green!70!black}{// FB loop}}
            \State{$\mathbf{X}^{j,0} = \mathbf{X}^{j-1}$}
            \State{\textbf{repeat for }{$i=1,2,\ldots,I$}}
            \State{\quad\quad$\hat{\mathbf{X}}^{j,i} = \mathbf{X}^{j,i-1} - \gamma \mathrm{Re} \left \{ \Phi^\dagger \left ( \Phi \left ( \mathbf{X}^{j,i-1} \right ) \right ) - \mathbf{X}^{\mathrm{dirty}} \right \} $}
            \State{\quad\quad $\mathbf{X}^{j,i} = \mathrm{prox}_{\gamma \lambda \widetilde{r}}\left (\hat{\mathbf{X}}^{j,i} \right ) $  \textcolor{green!70!black}{// Using Algorithm~\ref{algo:prox_hyperusara}}}
            \State{\textbf{until}$ \left . \left \|\mathbf{X}^{j,i}-\mathbf{X}^{j,i-1} \right \|_{\mathrm{F}} \middle / \left \|\mathbf{X}^{j,i-1} \right \|_{\mathrm{F}} \right . <\xi_2$}
        \State{$\mathbf{X}^{j} = \mathbf{X}^{j,i}$}
    \Until{$ \left . \left \|\mathbf{X}^{j}-\mathbf{X}^{j-1} \right \|_{\mathrm{F}} \middle / \left \| \mathbf{X}^{j-1} \right \|_{\mathrm{F}} \right . <\xi_1$}
    \State \Return {$\mathbf{X}^{j}$}
\end{ruledalgorithm}

\begin{ruledalgorithm}[algo:prox_hyperusara]{Solving proximity operator of \eqref{eq:sara_prior} with dual-FB}
    \Given{$\mathbf{X}^0$, $\mathbf{\hat{X}}$, $\bm{\Psi}$, $\bm{\Psi}^\dagger$, $\bm{w}$ \textcolor{green!70!black}{// Initialization}}
    \Given{$\gamma$, $\lambda$ \textcolor{green!70!black}{// Step size and regularization parameter}}
    \Given{$\xi_3$ \textcolor{green!70!black}{// Tolerance for dual-FB}}
    \Given{$K$ \textcolor{green!70!black}{// Maximum iterations for dual-FB}}
    \State{$\mathbf{V}^{0} = \Psi^\dagger \left ( \mathbf{X}^{0} \right ) $}
    \State{\textbf{repeat for }{$k=1,2,\ldots,K$}}
        \State{\quad\quad $\mathbf{X}^{k} = \mathrm{prox}_{\iota_{\mathbb{R}_+^{N \times L}} (\cdot)} \left ( \hat{\mathbf{X}} - \Psi \left( \mathbf{V}^{k-1} \right ) \right)$}
        \State{\quad\quad $\hat{\mathbf{V}}^{k} = \mathbf{V}^{k} + \Psi^\dagger \left ( \mathbf{X}^{k} \right )$}
        \State{\quad\quad $\mathbf{V}^{k} = \hat{\mathbf{V}}^{k} - \mathrm{prox}_{\gamma \lambda \| \cdot \|_{1,2,\bm{w}}} \left ( \hat{\mathbf{V}}^{k} \right ) $}
    \State{\textbf{until} $ \left . \left \| \mathbf{X}^{k}-\mathbf{X}^{k-1} \right \|_{\mathrm{F}} \middle / \left \| \mathbf{X}^{k-1} \right \|_{\mathrm{F}} \right . <\xi_3$}
    \State \Return {$\mathbf{X}^{k}$}
\end{ruledalgorithm}

%%%%%%%%%%%%%%%%%%%%%%%%%%%%%%%%%%%%%%%%%%%%%%
%%%%%%%%%%%%%%%%%%%%%%%%%%%%%%%%%%%%%%%%%%%%%%
%%%%%%%%%%%%%%%%%%%%%%%%%%%%%%%%%%%%%%%%%%%%%%
\section{HyperAIRI} \label{sec:hyperairi}

In this section, we present the HyperAIRI algorithm, which can be seen as the PnP counterpart of Hyper-uSARA within the PnP-FB. We subsequently detail the architecture of HyperAIRI denoisers and the associated training strategy, and we formalize the algorithmic structure of our proposed approach. 

\begin{figure*}[t]
    \centering
    \begin{minipage}[b]{0.530\linewidth}
        \centering
        \includegraphics[width=\linewidth]{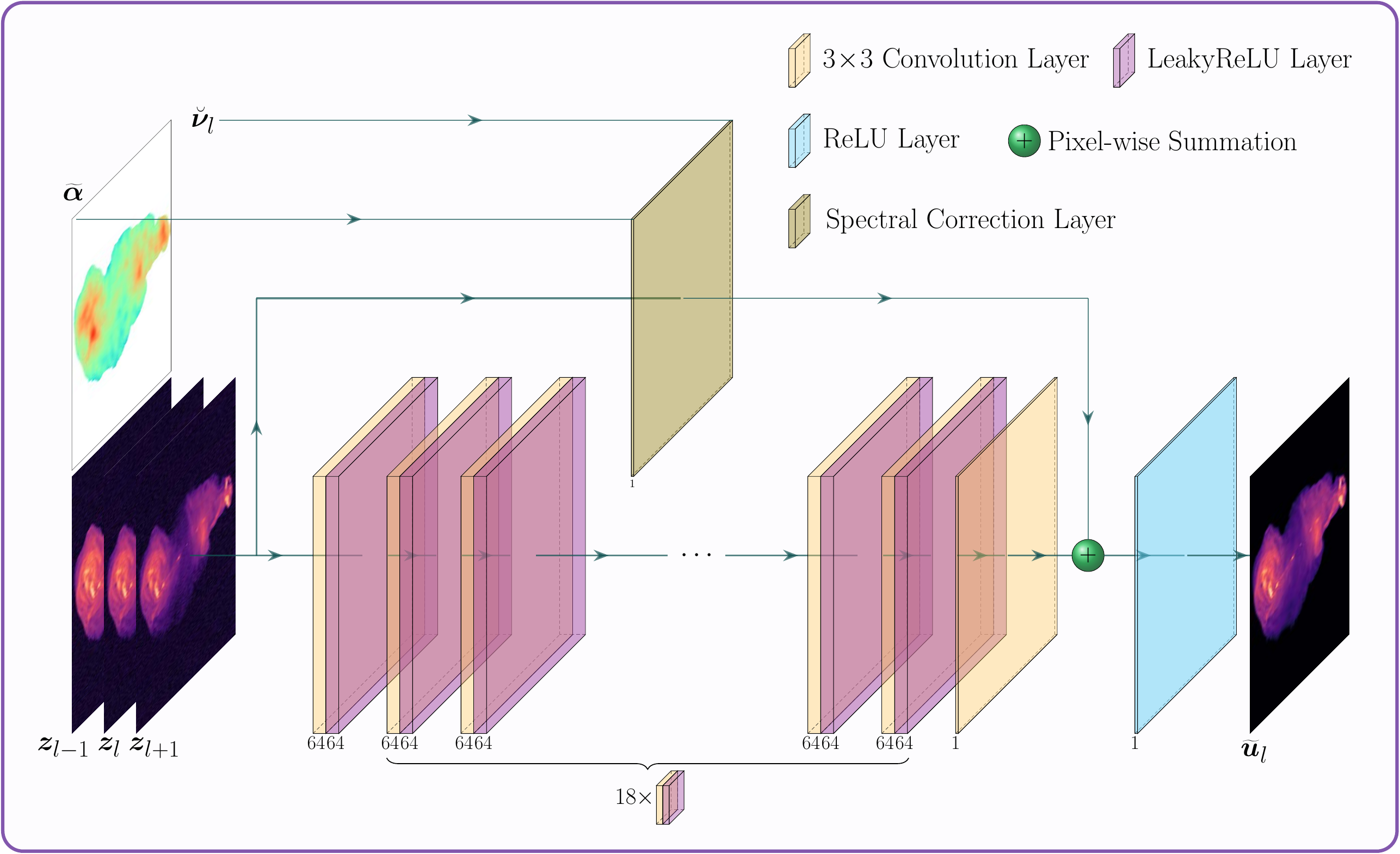}
        \par
        (a) HyperAIRI denoiser architecture
    \end{minipage}
    \begin{minipage}[b]{0.445\linewidth}
        \centering
        \includegraphics[width=\linewidth]{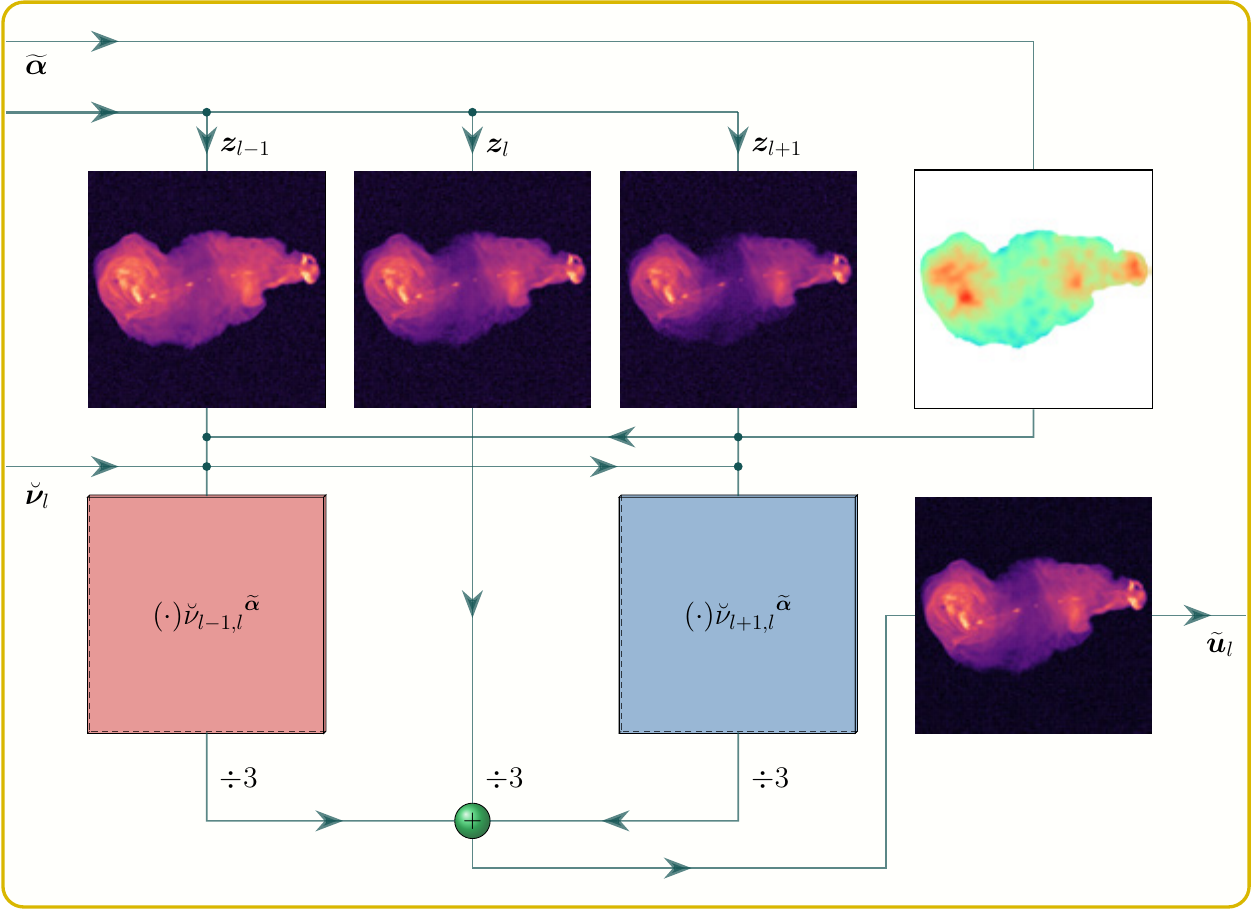}
        \par
        (b) spectral correction layer
    \end{minipage}
    \par\medskip
    \begin{minipage}[b]{0.228\linewidth}
        \centering
        \includegraphics[width=\linewidth]{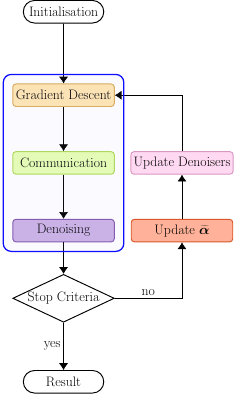}
        \par
        (c) overall flowchart
    \end{minipage}
    \begin{minipage}[b]{0.752\linewidth}
        \centering
        \includegraphics[width=\linewidth]{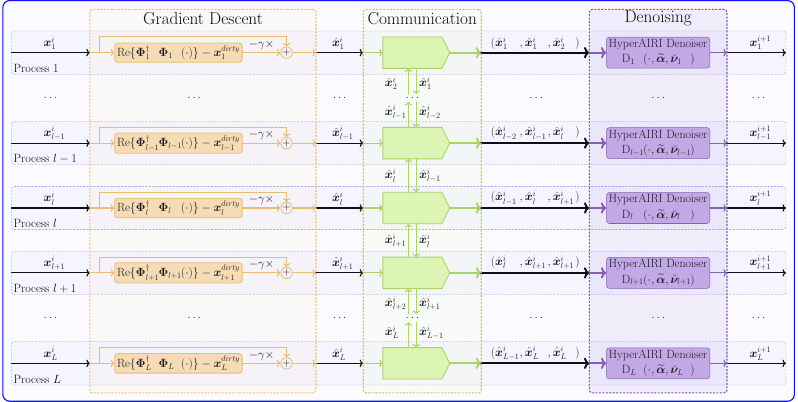}
        \par
        (d) FB iteration structure
    \end{minipage}
    
    \caption{HyperAIRI algorithm structure. Panels illustrate the following: (a) HyperAIRI denoiser architecture; (b) the inner structure of the spectral correction layer; (c) the flowchart of HyperAIRI algorithm; and (d) HyperAIRI's FB iteration structure.}
    \label{fig:ha_struct}
\end{figure*}

\subsection{Hyperspectral PnP-FB and HyperAIRI denoiser}

\added{From its} definition \eqref{eq:prox}, the proximal operator associated with a regularization function can be interpreted as solving a denoising problem. PnP algorithms build on this insight by replacing the proximal operator with a pre-trained DNN, typically trained for image denoising \citep{venkatakrishnan2013plug}. \added{PnP regularization is thus implicit, underpinned by the learned DNN denoiser. }
In the case of additive Gaussian noise, the denoising problem consists of estimating the ground truth $\overline{\mathbf{U}}$ from the noisy measurement $\mathbf{Z}$, modeled as
\begin{equation}
    \mathbf{Z} = \overline{\mathbf{U}} + \sigma\mathbf{E}
    \label{eq:train_noisy_cube}
\end{equation}%
where $\added{\sigma}\mathbf{E} \sim \mathcal{N}(0,\added{\sigma}\mathbf{I})$ is a realization of a white Gaussian noise cube with standard deviation $\sigma$. 
By construction, the trained denoisers are blind to the measurement process\added{, \textit{i.e.} the measurement operator model. Consequently}, they can adapt \added{seamlessly} %\added{without alteration} 
to RI imaging for different telescopes and under varying observational conditions. AIRI has been a successful instantiation of PnP algorithms for monochromatic RI imaging. By replacing the handcrafted proximity operator in FB with \added{a} learned firmly non-expansive denoiser\footnote{An operator $\mathrm{D}(\cdot)$ is firmly non-expansive if it satisfies the condition:  $\|\mathrm{D}(\bm{a}) - \mathrm{D}(\bm{b})\|^2 \leq \langle \mathrm{D}(\bm{a}) - \mathrm{D}(\bm{b}), \bm{a} - \bm{b} \rangle$. Intuitively, this means $\mathrm{D}(\cdot)$ moves inputs closer together while generally preserving their relative directions.}, \added{it leverages} the robustness of optimization algorithms and the flexibility of PnP algorithms. \added{The AIRI denoiser is \added{typically} trained for a well-defined noise level $\sigma$.  \citet{terris2023image} have shown that the effect of the noise level emulates that of the regularization parameter $\lambda$ in minimization tasks of the form~\eqref{eq:obj}, impacting significantly the efficiency of the underlying learned regularization. The authors also suggest that the adequate training noise level should correspond to the inverse of the target dynamic range of the final reconstruction.}

\added{We propose HyperAIRI, the PnP-variant of \eqref{eq:fb}, whose DNN denoiser denoted by $\mathrm{D}(\cdot)$ adapts AIRI's network architecture to accommodate multichannel input and exploit spectral correlations. HyperAIRI's iterative structure reads}
\begin{equation}
    \mathbf{X}^{i+1} = \mathrm{D} \left ( \mathbf{X}^i - \gamma \nabla f \left (\mathbf{X}^i \right ) \right ).
    \label{eq:pnp_fb}
\end{equation}%
\added{In hyperspectral RI imaging, the spectral dimension of the formed image cube can vary substantially, depending on the spectral behavior of the probed radio emission. We therefore adopt a sliding-window approach when applying the HyperAIRI denoiser to the image cube by sweeping along the spectral axis. This flexible approach decouples the number of DNN input channels from the spectral dimension of the target image cube. It also supports channel-wise parallel denoising with minimal cross-channel communication, since each denoiser utilizes only the correlations between its immediate neighbors. To mitigate the potential limitations of this local denoising, the power-law spectral model is encapsulated in the HyperAIRI denoiser to propagate global spectral information.}
 
\added{Specifically,} the HyperAIRI denoiser takes a DnCNN architecture with 20 convolutional layers. 
To denoise a channel $l$ of a noisy image cube $\mathbf{Z}$, the denoiser takes as input (i) the sub-cube composed of the target channel $\bm{z}_{l}$ and its two adjacent channels, (ii) the frequency ratios between the neighboring and target channels, and (iii) the estimated spectral index map.
It then outputs the denoised image $\widetilde{\bm{z}}_l$.  
The HyperAIRI denoiser embeds the power-law spectral model underpinning the hyperspectral image cube via a spectral correction layer shown in Figure~\ref{fig:ha_struct}(b).
This layer takes the same set of inputs of HyperAIRI denoisers, and provides as output the estimate of the target channel:
\begin{equation}
    \widetilde{\bm{u}}_l = \frac{1}{3} \left( \bm{z}_{l-1} {\breve{\nu}_{l-1,l}}^{\widetilde{\bm{\alpha}}}  + \bm{z}_{l} + \bm{z}_{l+1} {\breve{\nu}_{l+1,l}}^{\widetilde{\bm{\alpha}}} \right ),
\end{equation}%
where $\breve{\nu}_{l-1,l} = \nu_{l-1} / \nu_{l}$ and $\breve{\nu}_{l+1,l} = \nu_{l+1} / \nu_{l}$. In the spectral correction layer, only the spectral index map is involved, while the spectral curvature map is delegated implicitly to the other layers.
This separation supports the numerical stability of the entire algorithm, especially in cases where $\widetilde{\bm{\alpha}}$ is estimated dynamically during reconstruction.
The remaining layers in the DNN are designed to predict the residual difference between the initial estimate and the ground-truth image.
The final estimate of the HyperAIRI denoiser is the summation of the predicted difference and the output of the spectral correlation layer with positivity constraint from ReLU layer. The full architecture is shown in Figure~\ref{fig:ha_struct}(a).

\added{Choosing the sliding window to encompass three adjacent channels only meets our requirements for efficiency and flexibility. Firstly, this choice restricts border channels to only the first and last channels of the image cube. These two edge cases can be easily handled by duplicating them as the third input in their corresponding sub-cube. Secondly, with the noise varying significantly across channels due to the nature of the hyperspectral Fourier sampling, a small window size ensures that the underlying noise levels of the channels within it are sufficiently close to be treated as identical, which is consistent with the denoiser training. Thirdly, it allows the HyperAIRI denoiser to maintain a similar number of learnable parameters to AIRI, ensuring comparable training and inference computational costs.}

\subsection{Training HyperAIRI denoisers} \label{sec:train_ha}
 
To address the limited availability of training data for hyperspectral RI imaging, we build a training dataset based on the two monochromatic datasets proposed by \citet{terris2025airi}, namely the Optical Astronomical Image Dataset (OAID) and the Magnetic Resonance Image Dataset (MRID). Both datasets contain 5,000 images of size $512 \times 512$ pixels, and are used jointly during training.
For each monochromatic image, a corresponding hyperspectral image cube is generated according to the power-law model in \eqref{eq:power_law}. Following \citet{thouvenin2023parallelI}, the curvature map $\bm{\beta}$ is modeled as a random Gaussian field, while the spectral index map $\bm{\alpha}$ is derived from the logarithm of a Gaussian-blurred version of $\overline{\bm{u}}$ combined with a random Gaussian field ensuring spatial correlation.
The elements in $\bm{\alpha}$ are in the range $[-5,0]$ and those in $\bm{\beta}$ are in the range $[-0.5,0.5]$.
The simulated image cube $\overline{\mathbf{U}}\in\mathbb{R}^{N \times 3}$ is expressed as:
\begin{equation}
    \overline{\mathbf{U}} = \left ( \overline{\bm{u}} \breve{\nu}_{1,2} ^{ -\bm{\alpha} + \bm{\beta} \log \breve{\nu}_{1,2}  } , \overline{\bm{u}}, \overline{\bm{u}} \breve{\nu}_{3,2} ^{ -\bm{\alpha} + \bm{\beta} \log \breve{\nu}_{3,2} } \right ),
    \label{eq:train_cube}
\end{equation}%
where $\breve{\nu}_{1,2}$ and $\breve{\nu}_{3,2}$ are sampled uniformly from $[0.9,1.0]$ and $[1.0,1.1]$ respectively. The ratio 1.0 is included in the range to handle the boundary cases, where we duplicate the edge channels when denoising an image cube, resulting in frequency ratios of 1.0.

HyperAIRI denoisers are trained in a supervised manner using pairs of the simulated ground-truth image cubes $\overline{\mathbf{U}}$ and their noisy versions $ \mathbf{Z}$ obtained following \eqref{eq:train_noisy_cube}.
As indicated by \citet{terris2023image,terris2025airi}, ensuring the non-expansiveness of the denoisers is crucial for the convergence of the final PnP algorithm in reconstructing HDR RI images. Therefore, the loss function includes a standard $\ell_1$ loss and a Lipschitz regularization term. Denoting the denoiser as $\mathrm{D}$ with learnable parameters as $\bm{\theta}$, the training loss is
\begin{equation}
    \begin{aligned}
        \mathcal{L}(\bm{\theta}, \mathbf{Z},\mathbf{\overline{U}}, \bm{\alpha}, \breve{\bm{\nu}}_2) & = 
        \| \operatorname{D}_{\bm{\theta}}(\mathbf{Z},\bm{\alpha},\breve{\bm{\nu}}_2)-(\overline{\mathbf{U}})_2\|_1 \\
         + & \kappa \, \mathrm{max} \{ \| \boldsymbol{\nabla} \operatorname{Q}_{\bm{\theta}}(\overline{\mathbf{U}})\|_{\rm{S}}, 1-\delta \},
    \end{aligned}
    \label{eq:train_loss}
\end{equation}%
where $(\cdot)_2$ selects the second channel of the image cube, $\bm{\alpha}$ and $\breve{\bm{\nu}}_2 = (\breve{\nu}_{1,2}, \breve{\nu}_{3,2})$ are the spectral index map and frequency ratios for $\mathbf{U}$. In the regularization term, $\mathrm{Q}$ is defined as $2\mathrm{D}_{\bm{\theta}}(\cdot,\bm{\alpha},\breve{\bm{\nu}}_2) - (\cdot)_2$, and $\bm{\nabla} \operatorname{Q} (\cdot)$ denotes the Jacobian of $\operatorname{Q}$. Here, $\kappa > 0$ is the regularization parameter, and $\delta > 0$ determines the lower bound of the regularization term.

To avoid retraining for each specific problem, a shelf of HyperAIRI denoisers can be trained across various $\sigma$ values.
In this work, we trained 6 DNNs at noise levels in the set \added{$\{ 1, 2, 4, 8, 16, 32 \} \times 10^{-5}$, corresponding to a target dynamic range for the reconstructed image cube in the interval $[10^3,~ 10^5]$, which is consistent with the expected dynamic range enabled by modern telescopes. During reconstruction, an appropriate denoiser is selected from the shelf based on the estimate of the target dynamic range, and a scaling factor is applied to the input of the denoiser to match its noise level. More details are provided in Section~\ref{sec:algo_structure}.}

Similarly to AIRI denoisers, HyperAIRI denoisers are trained using randomly cropped patches of size $46 \times 46 \times 3$ from the simulated image cubes to speed up the training process owing to their convolutional nature. The compact receptive field allows application to any image size during image reconstruction.

The training scripts are built with PyTorch in Python.
\added{Given the similarity between the network architectures of AIRI and HyperAIRI denoisers, we follow the same training settings as in \citet{terris2025airi}.}
Each HyperAIRI denoiser is trained for 5000 epochs with the ADAM optimizer \citep{kingma2014adam}. The initial learning rate is set to $10^{-4}$, $10^{-5}$ and $10^{-6}$ for training noise levels \added{$\{32\} \times10^{-5}$, $\{4, 8, 16\}\times10^{-5}$ and $\{1, 2\}\times10^{-5}$} respectively. The learning rate is halved every 900 epochs. The regularization factor $\kappa$ in \eqref{eq:train_loss} is also fine-tuned for each DNN to make sure the final value of the regularization term is as close to 1 as possible. 
In practice, $\kappa$ is typically close to the average value of the $\ell_1$ loss term for optimal results. 
As suggested by \citet{terris2023image}, we also use the weights of the DNNs trained with $\kappa=0$ as initialization to improve the training stability and the performance of the final DNNs.

\subsection{Algorithmic structure} \label{sec:algo_structure}

Plugging HyperAIRI denoisers trained at desired noise levels into the iterative FB forms the HyperAIRI algorithm. The detailed structure of our proposed algorithm is shown in Figure~\ref{fig:ha_struct}.(c). 

\subsubsection{Initialization} \label{sec:algo_init}

In the initialization stage,  \added{the adequate} denoiser for each channel is selected from the pre-trained denoiser shelf on \added{the basis of the estimate of} its dynamic range. 
As proposed by \citet{terris2023image}, \added{the target dynamic range at channel $l$ can be estimated as $\widetilde{\rho}_l /\sigma_{\mathrm{heu}, l}  $}, where $\sigma_{\mathrm{heu},l}$ is the estimate of the image-domain noise level given in \eqref{eq:heuristic}, and $\widetilde{\rho}_l$ is \added{the peak value of the reconstructed} image at channel $l$. 
An initial estimate of $\widetilde{\rho}_l$ is obtained from the peak value of the dirty image normalized by the peak value of the point spread function (PSF), such that
\begin{equation}
    \widetilde{\rho}_l = \left . \max \left ( \bm{x}_l^{\mathrm{dirty}} \right ) \middle / \max \left ( \mathrm{Re} \{ \bm{\Phi}_l^\dagger \bm{\Phi}_l \bm{\delta} \} \right ) \right . ,
    \label{eq:rho}
\end{equation}%
where $\bm{\delta} \in \mathbb{R}^N$ is a Dirac delta image with value 1 at its center and 0 otherwise. For a given channel $l$, the denoiser with a noise level $\sigma$ closest to but smaller \added{than the inverse of} the target dynamic range estimate $\sigma_{\mathrm{heu},l} / \widetilde{\rho}_l$ is selected. To \added{match} the noise level of the selected denoiser, a scaling factor is applied \added{to its input sub-cube. Formally, the channel-wise denoising operator $\operatorname{D}_l$ is given by}
\begin{equation}
    \operatorname{D}_l(\cdot, \widetilde{\bm{\alpha}}, \breve{\bm{\nu}}_l) = \varrho_l \operatorname{D}_{\sigma}(\cdot / \varrho_l, \widetilde{\bm{\alpha}}, \breve{\bm{\nu}}_l),
\end{equation}%
with $\varrho_l = \sigma_{\mathrm{heu},l} / \sigma$ \citep{dabbech2022first}.
Finally, the spectral index map $\widetilde{\bm{\alpha}}$ can be initialized to $\bm{0}$, derived from the dirty image, or specified by the user. 

\subsubsection{\added{HyperAIRI iteration structure}}

\added{The HyperAIRI iteration structure based on FB is summarized in Figure~\ref{fig:ha_struct}(d)}. At iteration $i$, the process assigned to each channel $l$ \added{performs the forward step by evaluating the gradient of the data-fidelity term at $\bm{x}_l^{i}$, identified as the negative residual dirty image $\bm{x}_l^{\mathrm{res}, i}$, and applies a gradient-descent update to obtain $\hat{\bm{x}}_l^{i}$, as $\hat{\bm{x}}_l^{i}=\bm{x}_l^{i} + \gamma \bm{x}_l^{\mathrm{res}, i}$. To conduct the backward step, the process then communicates $\hat{\bm{x}}^i_l$} to the adjacent channels ($l-1$ and $l+1$) and receives their corresponding images. These three images are combined to form the input \added{sub-cube} $\hat{\mathbf{X}}^i_l = (\hat{\bm{x}}^i_{l-1}, \hat{\bm{x}}^i_l, \hat{\bm{x}}^i_{l+1})$, which is then \added{denoised} by the denoiser selected for channel $l$ \added{yielding the updated image $\bm{x}^{i+1}_l$}.
Specific to edge channels, the processes for the first and last channels form image \added{sub-cubes} by communicating with their only neighbor, resulting in $\hat{\mathbf{X}}^i_1 = (\hat{\bm{x}}^i_{1}, \hat{\bm{x}}^i_1,\hat{\bm{x}}^i_{2})$ and $\hat{\mathbf{X}}^i_L = (\hat{\bm{x}}^i_{L-1}, \hat{\bm{x}}^i_L,\hat{\bm{x}}^i_{L})$, respectively.
Thereafter, the denoisers for each channel are applied simultaneously to \added{all sub-cubes $\hat{\mathbf{X}}^i_l$}. 
To exploit the transformation invariance of images, random $90^\circ$ rotations and flips, and corresponding inverse transformations, are applied to the inputs and outputs of the DNNs, respectively \citep{terris2023image}.
The \added{denoising} result of the \added{backward} step \added{from} all channels forms the image cube update $\mathbf{X}^{i+1}$.

\subsubsection{Dynamic denoiser selection and spectral index map update} \label{sec:algo_adap}
\added{Inspired by \citet{terris2025airi},  HyperAIRI allows for the dynamic selection of the adequate denoisers from the shelf within the FB iterations. 
Often, the initial value of $\widetilde{\rho}_l$ obtained from \eqref{eq:rho} tends to be overestimated, leading to a suboptimal denoiser selection. To address this issue, at each iteration, we update $\widetilde{\rho}_l$ as the peak value of ${\bm{x}}^{i+1}_l$. The ratio $\sigma_{\mathrm{heu},l}/\widetilde{\rho}_l$} is then monitored to decide whether the current denoiser $\operatorname{D}_\sigma$ and its corresponding scaling factor $\varrho_l$ should be updated.

We adopt a similar strategy to update the estimate of the spectral index map $\widetilde{\bm{\alpha}}$ from the current estimate of the cube $\mathbf{X}^{i}$, when required. Ignoring the spectral curvature term in \eqref{eq:power_law}, setting $\nu_0 = 1$ for simplicity and applying logarithm to both sides, the pixel $n$ in channel $l$ should follow
\begin{equation}
    \log{\bm{x}_{l}^i} = -\widetilde{\bm{\alpha}} \log{\nu_l} + \log{\bm{x}_{0}^{i}}.
    \label{eq:power_law_log}
\end{equation}% 
The slope of this linear \added{mapping between $\log{\bm{x}_{l}}$ and $\log{\nu_l}$} can be evaluated pixel-wise using linear regression \added{to obtain the} estimate of the spectral index map  $\widetilde{\bm{\alpha}}$.
In practice, a smoothing 2D Gaussian kernel is applied to the image cube before \added{applying} linear regression \added{for} numerical stability.
The 2D Gaussian kernel is fitted to the PSF's primary lobe at the highest-frequency channel, which represents the highest spatial resolution of the measurements.
Additionally, pixels with intensities below a preset threshold, taken as the minimal heuristic noise level of the entire image cube, are excluded from the estimation of $\widetilde{\bm{\alpha}}$.

\subsubsection{Fully parallelized implementation}

HyperAIRI reconstruction is executed in a distributed, channel-wise fashion, together with an efficient large-scale RI measurement operator implementation \citep{dabbech2025distributed}.
During initialization, each process is assigned its channel's back-projected dirty image, local measurement operator, selected denoiser and initial spectral-index map.
At each FB iteration, gradient-descent and denoising steps then run concurrently across channels\added{, as in Figure~\ref{fig:ha_struct}(d)}. Each channel process exchanges only its immediate neighbors' images to assemble the three-slice input cubes between these two steps, minimizing communication overhead.
As with the AIRI denoiser, the HyperAIRI denoiser has spatial faceting functionality. The denoisers can be applied on the overlapped facets extracted from $\hat{\mathbf{X}}^i_l$ in parallel.
This hybrid spatial-spectral parallelization scales nearly linearly with both the number of channels and available compute resources.
As a result, our implementation delivers a highly efficient solver capable of handling large-scale hyperspectral datasets in practical runtimes.

%%%%%%%%%%%%%%%%%%%%%%%%%%%%%%%%%%%%%%%%%%%%%%
%%%%%%%%%%%%%%%%%%%%%%%%%%%%%%%%%%%%%%%%%%%%%%
%%%%%%%%%%%%%%%%%%%%%%%%%%%%%%%%%%%%%%%%%%%%%%

\section{Validation on simulated VLA observations}
\label{sec:sim_exp}

In this section, we study the performance of different algorithms for reconstructing hyperspectral RI image cubes from simulated observations.

\subsection{Test dataset}

\begin{figure}
    \centering
    \begin{tabular}{c @{\hspace{0.8\tabcolsep}} c}
        \includegraphics[width=0.45\linewidth]{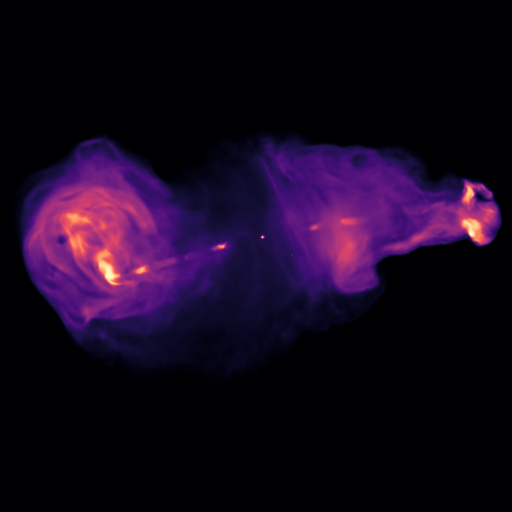}
        &
        \includegraphics[width=0.45\linewidth]{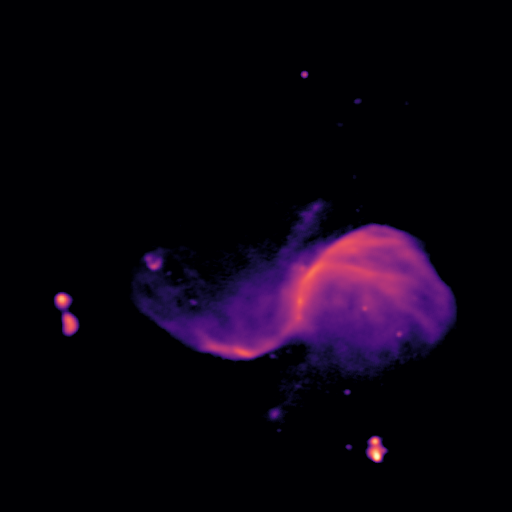}\\
        (a) 3C~353 & (b) Messier~106\\
        \includegraphics[width=0.45\linewidth]{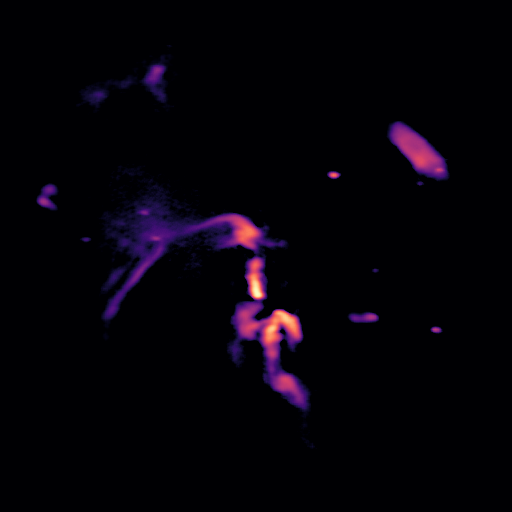}
        &
        \includegraphics[width=0.45\linewidth]{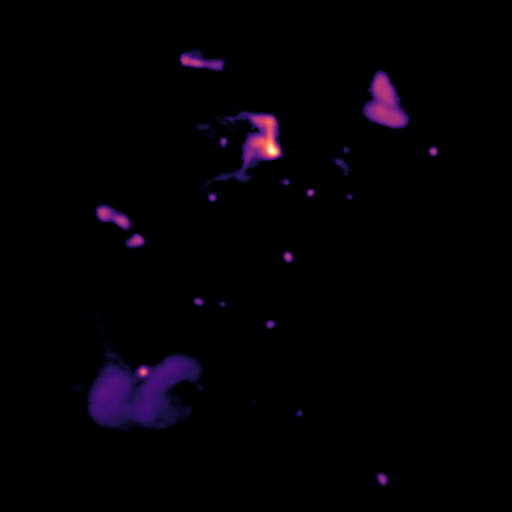}\\
        (c) Abell~2034 & (d) PSZ2~G165.68+44.01
    \end{tabular}
    \caption{Original ground-truth images for generating the simulated test dataset. Each image has a size of $512 \times 512$ and the names of the sources are listed below the images.}
    \label{fig:sim_gdth}
\end{figure}

The considered ground-truth images are derived from four radio images of size $N = 512\times512$, which consist of giant radio galaxies 3C~353 \citep{swain1998internal} and Messier~106 \citep{shimwell2022vizier}, and the radio galaxy clusters Abell~2034 and PSZ2~G165.68+44.01 \citep{botteon2022planck} shown in Figure~\ref{fig:sim_gdth}.
The images are first preprocessed to reduce noise and artifacts, then randomly exponentiated to make the final dynamic range in the interval $[10^3,10^5]$. We follow the same preprocessing strategies for OAID and MRID and create 40 HDR monochromatic images. 
Following \citet{thouvenin2023parallelI}, for each image, a spectral index map and a curvature map are generated, in a manner consistent with the procedure used for the training dataset. Specifically, the spectral index map is modeled as the sum of a blurred version of the ground truth in logarithmic scale and a Gaussian field, clipped to the range $[-5,0]$. The curvature map is modeled as a Gaussian field, clipped to the range $[-0.5, 0.5]$. 
Each image cube contains $L=36$ channels, with the first channel fixed at the observation frequency of 1 GHz and the last channel randomly chosen from the interval $[1.5,2.0]$ GHz. The observation frequencies of the remaining channels are uniformly spaced within this range.
We simulate 40 ground-truth image cubes of size $512 \times 512 \times 36$ following the spectral model \eqref{eq:power_law}.

Hyperspectral RI data are simulated using the antenna configurations of the VLA. The telescope is known for its unique ability to change the positions of the antennas along rail tracks, forming four standard configurations with maximum baselines ranging from 36.4 km to 1.03 km. Configurations with larger baselines provide higher angular resolution, while those with shorter baselines offer higher sensitivity to spatially extended emission.
Astronomers often combine measurements from different configurations to obtain a comprehensive Fourier sampling of the target object. Under these considerations, we simulate RI observations combining configurations A and C of the VLA. The Fourier sampling patterns for both configurations are generated using the MeqTrees software package \citep{noordam2010meqtrees}.
We randomly vary the telescope's pointing direction, and the observation duration within the range $[5,10]$ hours, and fix the integration time to 12 seconds. The resulting total number of points in each sampling pattern, at any given spectral channel, varies from $1.1\times10^6$ to $2.1\times10^6$. Examples of simulated sampling patterns for spectral channels with different observation frequencies are shown in Figure~\ref{fig:sim_uv}. 
The hyperspectral RI data are simulated under the assumption of a narrow field of view, and following \eqref{eq:model_wideband}, where the standard deviation of the observation noise is fixed across the spectral channels aligning with the target dynamic range of the cube as per the heuristic \eqref{eq:heuristic}, such that the target dynamic range of the reconstructed image cube matches that of the corresponding ground-truth image cube. The observation's nominal resolution is parameterized by the super-resolution factor, defined as the ratio between the image spatial Fourier bandwidth and the maximum projected baseline. We fix the super-resolution factor at the highest-frequency channel to 1.5, leading to a super-resolution factor across different channels in the range $[1.5, 3.0]$. 
The resulting test dataset is composed of 40 inverse problems.

\subsection{Benchmark settings}

For benchmarking, we compare HyperAIRI's performance against four algorithms. These are its optimization-counterpart Hyper-uSARA, the hyperspectral CLEAN variant in WSClean \citep{offringa2017optimized}, and the monochromatic algorithms AIRI and uSARA which are applied for each channel independently. Briggs weighting with robustness parameter set to 0 is considered for all algorithms. 

In our experiments, the regularization parameters for the optimization algorithms and the effective noise levels for PnP algorithms are directly linked to the heuristic noise level via \eqref{eq:heuristic} without further tuning.
WSClean parameters are tuned on a small subset of the test dataset and then fixed across the entire dataset to ensure the best overall image quality.
In particular, the auto-masking and auto-thresholding values are set to 1.0 and 0.5, respectively, and the number of polynomial terms for spectrum fitting is fixed at 4. The full WSClean command is shown in Appendix~\ref{sec:clean_cmd}. 
We also note that the WSClean reconstructions are normalized by the area of the CLEAN beam $A_{\mathrm{beam}}$ on a per-channel basis, thus converting the unit of pixel intensities from Jy/beam to Jy/pixel to align with the ground-truth image cube. The normalization factor is given by
\begin{equation}
    A_{\mathrm{beam}} = \frac{\pi B_{\mathrm{maj}} B_{\mathrm{min}} }{4 \log{2}},
    \label{eq:clean_beam_area}
\end{equation}%
where the respective $B_\mathrm{maj}$ and $B_{\mathrm{min}}$ are the size of the CLEAN restoring beam in the major and minor axes in the unit of pixel.

HyperAIRI, Hyper-uSARA, AIRI and uSARA are \added{implemented in MATLAB within a unified modular framework for large-scale RI imaging, underpinned by a fully distributed widefield RI measurement operator~\citep{dabbech2022first,dabbech2025distributed}. We recall that WSClean is a C++ software}.
All experiments were run on \href{https://www.epcc.ed.ac.uk/hpc-services/cirrus}{Cirrus}, a UK Tier 2 HPC service that provides access to both CPU and GPU compute nodes. Each CPU node comprises two Intel 18-core Xeon E5-2695 processors with 256 GB of memory, while each GPU node comprises two 20-core Intel Xeon Gold 6148 processors, four NVIDIA Tesla V100-SXM2-16GB GPUs, and 384 GB of memory. For each inverse problem in the test dataset, uSARA, WSClean and Hyper-uSARA run on a single CPU node, whereas AIRI and HyperAIRI were executed on a single GPU node to accelerate the inference of DNNs. Each AIRI and HyperAIRI denoisers is trained with half GPU node.

\begin{table*}[t]
    \centering
    \caption{Numerical results of the various algorithms.
    Reconstruction quality of the image cube is reported in terms of SNR, SNR of the logarithmic scale image cubes (logSNR), residual-to-dirty ratio (RDR) and the SNR of the spectral index maps (sSNR).
    Computational performance details include the total number of iterations, the number of CPU cores and GPUs used in each run, the average timings for each reconstruction as well as for individual gradient (forward) and denoising (backward) steps. Values are presented in the format ``average $\pm$ standard deviation'' except for computing resources. The average RDR of the ground truth images is $(2.46 \pm 1.37) \times 10^{-3}$.
    Note that for WSClean, the number of iterations indicates the number of major cycles.
    }
    \setlength{\tabcolsep}{3pt}
    \begin{adjustbox}{max width=\textwidth,center}
        \hspace*{-0.175\textwidth}
        \begin{tabular}{lccccccccccc}
            \toprule
            \multirow{2}{*}{Algorithm} & Joint & SNR & logSNR & RDR & sSNR & Iterations & \multicolumn{2}{c}{{Resources}} & {${t}_{\text{total}}$} & {${t}_{\text{gradient}}$} & {${t}_{\text{denoising}}$} \\
            & deconv. & {$\pm$ std (dB)} & {$\pm$ std (dB)} & {$\pm$ std $ ( \times 10^{-3} ) $} & {$\pm$ std (dB)} & {$\pm$ std} & {CPUs} & {GPUs} & {$\pm$ std (hr)} & {$\pm$ std (s)} & {$\pm$ std (s)} \\
            \midrule
            uSARA & No & 28.32 \(\pm\) 3.24 & 21.17 \(\pm\) 3.56 & 2.27 \(\pm\) 1.32 & 9.54 \(\pm\) 2.34 &  1251 \(\pm\) 299 & 36 & 0 & 
            2.60 \(\pm\) 0.55 &
            2.10 \(\pm\) 0.27 & 5.24 \(\pm\) 6.73 \\
            AIRI & No & 28.70 \(\pm\) 2.36 & 20.67 \(\pm\) 5.01 & 2.62 \(\pm\) 1.41 & 9.18 \(\pm\) 3.90 & 3000 \(\pm\) 0 & 36 & 4 & 
            1.84 \(\pm\) 0.14 & 
            1.60 \(\pm\) 0.23 & 0.46 \(\pm\) 0.32 \\
            \midrule
            WSClean & Yes & 11.27 \(\pm\) 1.57 & 12.05 \(\pm\) 4.08 & 2.36 \(\pm\) 1.27 & 3.81 \(\pm\) 5.91 & 10 \(\pm\) 1 & 36 & 0 & 
            \textbf{0.56 \(\pm\) 0.24} &
            -- & -- \\
            Hyper-uSARA & Yes & 29.30 \(\pm\) 2.91 & 22.30 \(\pm\) 4.15 & 2.32 \(\pm\) 1.34 & \textbf{11.35 \(\pm\) 3.78} & 1827 \(\pm\) 675 & 36 & 0 & 
            2.42 \(\pm\) 0.87 &
            2.09 \(\pm\) 0.27 & 2.58 \(\pm\) 1.43 \\
            HyperAIRI & Yes & \textbf{30.53 \(\pm\) 1.87} & \textbf{23.07 \(\pm\) 5.13} & \textbf{2.44 \(\pm\) 1.36} & 10.87 \(\pm\) 3.91 & 3000 \(\pm\) 0 & 36 & 4 & 
            1.90 \(\pm\) 0.14 &
            1.61 \(\pm\) 0.23 & 0.52 \(\pm\) 0.25 \\
            \bottomrule
        \end{tabular}
    \end{adjustbox}
    \label{tab:sim_metrics}
\end{table*}

\subsection{Evaluation metrics}

To quantitatively compare the reconstruction quality of the various algorithms in terms of accuracy to the ground truth, we use two metrics. The first one is signal-to-noise (SNR) ratio. Given the ground truth $\overline{\mathbf{X}}$ and an estimate image cube ${\mathbf{X}}$, the SNR is calculated as
\begin{equation}
    \mathrm{SNR}({\mathbf{X}}, \overline{\mathbf{X}}) = 20 \log_{10} \left ( \frac{\|\overline{\mathbf{X}}\|_\mathrm{F}}{\|\overline{\mathbf{X}} - {\mathbf{X}}\|_\mathrm{F}} \right ).
    \label{eq:snr_cube}
\end{equation}%
We also evaluate the SNR at each channel $l$ which reads
\begin{equation}
    \mathrm{SNR}({\bm{x}_l}, \overline{\bm{x}}_l) = 20 \log_{10} \left ( \frac{\|\overline{\bm{x}}_l\|_2}{\|\overline{\bm{x}}_l - \bm{x}_l\|_2} \right ).
    \label{eq:snr_channel}
\end{equation}%
To evaluate the algorithms' performance in reconstructing faint emissions, we evaluate the SNR on logarithmic scale images, denoted as logSNR.
Given a ground-truth image cube, with a dynamic range $a$, we apply the logarithmic scale transform parameterized by $a$ such that
\begin{equation}
    \mathrm{rlog}(\mathbf{X}) = \max({\mathbf{X}}) \log_a \left( \frac{a}{\max({\mathbf{X}})} \mathbf{X} + 1\right),
\end{equation}%
where $\max(\cdot)$ extract the maximum pixel intensities in the input image cube.
The overall and channel-wise SNRs are then computed on these logarithmically transformed image cubes with \eqref{eq:snr_cube} and \eqref{eq:snr_channel} as $\mathrm{SNR}(\mathrm{rlog}({\mathbf{X}}), \mathrm{rlog}(\overline{\mathbf{X}}))$ and $\mathrm{SNR}((\mathrm{rlog}({\mathbf{X}}))_l, (\mathrm{rlog}(\overline{\mathbf{X}}))_l)$, yielding the overall and channel-wise logSNR metrics respectively.

In assessing data fidelity in the image domain, we introduce the residual-to-dirty ratio (RDR). \added{The} overall and channel-wise RDR values can be calculated as 
\begin{equation}
   \mathrm{RDR}(\mathbf{X}^{\mathrm{res}}, \mathbf{X}^{\mathrm{dirty}}) =  \| \mathbf{X}^{\mathrm{res}} \|_{\mathrm{F}} / \| \mathbf{X}^{\mathrm{dirty}} \|_{\mathrm{F}}
\end{equation}%
and 
\begin{equation}
     \mathrm{RDR}(\bm{x}^{\mathrm{res}}_l, \bm{x}^{\mathrm{dirty}}_l) = \| \bm{x}^{\mathrm{res}}_l \|_2 / \| \bm{x}^{\mathrm{dirty}}_l \|_2 
\end{equation}%
respectively. If a reconstruction's RDR closely matches that of the corresponding ground-truth image, the reconstruction demonstrates high data fidelity.

Finally, the spectral index map from a given image cube $\mathbf{X}$, is obtained via linear regression applied at each pixel position to estimate the slope between the logarithm of spectral frequencies and the pixel intensities along the spectral axis following \eqref{eq:power_law_log}. 
To ensure numerical stability, the spectral index is only evaluated at pixels whose values above heuristic noise levels \eqref{eq:heuristic} across all the channels. Consider the ground-truth spectral index map $ \overline{\bm{\alpha}}$, the accuracy of the reconstructed spectral index map $\bm{\alpha}$ is measured by the spectral-index SNR metric (sSNR), defined as $\mathrm{SNR(\bm{\alpha, \overline{\bm{\alpha}}})}$.

\subsection{Reconstruction results}

\begin{figure}
    \centering
    \includegraphics[width=0.98\linewidth]{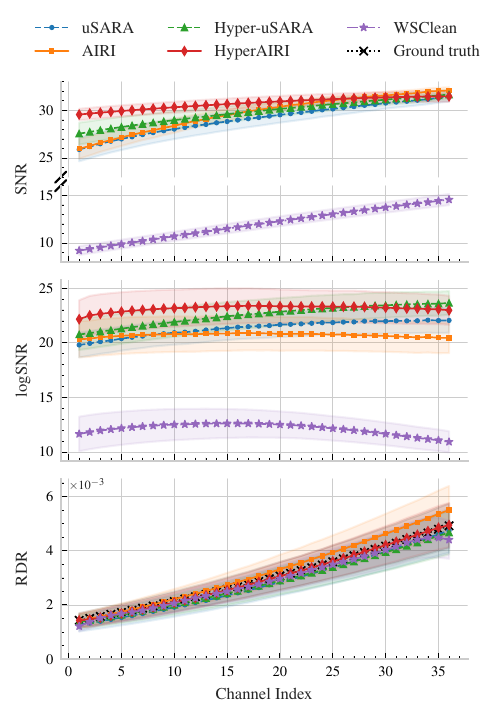}
    \caption{Channel-wise reconstruction metrics for various methods on the simulated test set. 
    The channel index ranges from 1 to 36. From top to bottom, the panels show SNR, logSNR and RDR versus channel index, respectively. Each point represents the average metrics value for the corresponding channel index across all problems, and the shaded areas indicate the $95\%$ confidence interval.}
    \label{fig:metrics_ch}
\end{figure}
\input{fig_sim_3c353}

Firstly, we discuss the overall numerical performance of the various methods on the simulated test dataset, as reported in Table~\ref{tab:sim_metrics}, where metrics are computed over the entire image cubes. 
We also examine channel-wise metrics illustrated in Figure~\ref{fig:metrics_ch}, where each point represents the average across the test cases, and the shaded region denotes $95\%$ confidence interval.

To analyze the fidelity of the reconstructed image cubes to the ground truth, we first examine the SNR metric quantifying how faithfully bright features are recovered. The hyperspectral algorithms Hyper-uSARA and HyperAIRI improve the mean SNR over their respective monochromatic counterparts uSARA and AIRI by about 1 and 1.5 dB. As expected, WSClean's reconstruction quality ranks lowest, a consequence of smoothing with a channel-specific CLEAN beam and adding with the residual dirty image. Figure \ref{fig:metrics_ch} shows that, generally, channel-wise SNR obtained by all algorithms increases with the frequency. This behavior could be explained by the nature of the hyperspectral Fourier sampling, where larger spatial bandwidths are probed at the highest frequencies. HyperAIRI exploits inter-channel correlations effectively, drastically narrowing the gap in SNR among the low and high frequencies. Similar behavior is observed with Hyper-uSARA although to a lesser extent. We then examine the logSNR metric results, assessing the recovery of faint emissions, shows that Hyper-uSARA and HyperAIRI exceed their monochromatic counterparts in overall average logSNR by 1.1 and 2.4 dB respectively, with WSClean again lagging behind in quality. The channel-wise logSNR curves in Figure~\ref{fig:metrics_ch} highlights the overall superior performance of HyperAIRI. Furthermore, we notice a consistent performance of the PnP approaches across channels, with a minor drop in the logSNR values at both ends, whereas the optimization-based algorithms exhibit logSNR values increasing with the frequency. In contrast, WSClean displays more pronounced drops in logSNR at the lowest and highest frequencies.

Assessing reconstruction data consistency in the image domain via the RDR metric indicates that all methods achieve good performance, with average RDR values close to the ground-truth reference.
uSARA falls slightly below the reference, suggesting minor over-fitting, whereas AIRI lies just above, indicating slight under-fitting. Figure~\ref{fig:metrics_ch} shows that HyperAIRI's channel-specific RDR curve follows the ground truth across all channels. Overall, HyperAIRI provides the best data fidelity both in aggregate and on a per-channel basis.

We examine the spectral accuracy of the reconstructed images via the sSNR metric. The hyperspectral algorithms Hyper-uSARA and HyperAIRI improve over their monochromatic counterparts, confirming the benefit of exploiting spectral correlations. Specifically, Hyper-uSARA achieves the highest average sSNR, closely followed by HyperAIRI. The slightly lower sSNR achieved by the latter may be related to its relatively small receptive field in both spatial and spectral dimensions. In contrast, WSClean exhibits significantly lower sSNR, likely due to its inherent high-sensitivity to variations in spatial resolution across different channels.

Secondly, we conduct a qualitative analysis, by analyzing the reconstructed images of a selected inverse problem, whose ground-truth cube is derived from the image of the radio galaxy 3C~353. Figure~\ref{fig:sim_recons} shows the reconstructions of the first and last channels obtained with different algorithms. 
The visual results are generally consistent with the quantitative analysis.
By design, WSClean exhibits lower resolution and dynamic range.
For the optimization-based algorithms, Hyper-uSARA exhibits fewer wavelet-like artifacts around the edges of bright structures than its monochromatic counterpart (see the two zoom regions), particularly in the first channel. As for the PnP algorithms, they both exhibit fewer ringing artifacts than uSARA and Hyper-uSARA, especially around the central black hole and the bright emission in the top-right zoom. Furthermore, HyperAIRI produces sharper depiction of relatively bright structures.
For instance, the central point source is more compact in the HyperAIRI reconstruction, with a more homogeneous radius across spectral channels.
However, compared to Hyper-uSARA, HyperAIRI appears to capture a smooth depiction of the faint extended emission whose intensity values are 3 to 4 orders of magnitude lower than the peak pixel value.
This trend, particularly noticeable at the highest frequency, is also reflected in the channel-wise logSNR metric. This could be due to the effective noise level and the resulting choice of the denoiser.
Lowering the heuristic noise level in the highest-frequency channel may help restore those faint features in this case. 

The inspection of the residual dirty images shows that WSClean exhibits good data fidelity in the image domain, with the signature of the radio emission being hardly noticeable.
As for Hyper-uSARA, while improving over its monochromatic counterpart, still shows signs of data over-fitting. 
Finally, HyperAIRI achieves comparable data fidelity to that of WSClean. Its residual dirty images are generally homogeneous, particularly in comparison to AIRI, apart from some residual structure in the first channel at the pixel positions of the two hotspots.

The spectral index maps produced by all methods generally agree with the ground truth.
Upon closer inspection, the spectral index maps of WSClean exhibit fake details which strongly correlate with image spatial features (see the two zooms).
This effect arises from CLEAN's high sensitivity to the varying spatial resolution across channels.
In contrast, the spectral index map of both Hyper-uSARA and HyperAIRI are more faithful to the ground truth, with both algorithms substantially reducing the artifacts present in their monochromatic counterparts.
When comparing these two hyperspectral algorithms, the spectral index maps of HyperAIRI shows even milder ringing artifacts (see the zooms of the central black hole and the right hotspot) but appears marginally more grainy in the bottom region, which is linked to the smoothness of its reconstruction in the higher-frequency channels.

The computational details of various algorithms are indicated in Table~\ref{tab:sim_metrics}. 
As expected, WSClean appears to be 5 and 4 times faster than optimization algorithms and PnP algorithms respectively. Its computational efficiency is explained by its small number of passes through the measurement operators within its major cycle.
In comparison, the rest of the algorithms, sharing the same iterative structure, are highly iterative in nature.
On average, both optimization-based algorithms require comparable computational time, despite Hyper-uSARA demanding nearly $50\%$ more iterations than uSARA. This highlights the efficiency of the hyperspectral regularization, whose sub-iterative proximity operator converges at least twice as fast. PnP algorithms also exhibit comparable computational time. Although HyperAIRI denoisers are more complex than AIRI, they do not introduce additional computational cost.
Notably, PnP approaches generally require more iterations than pure optimization methods. Nevertheless, they remain faster overall due to the efficient GPU-based inference of their DNN denoisers. In addition, the gradient steps of the PnP algorithms are slightly faster than those of the optimization-based algorithms, primarily because of differences in CPU performance and memory configuration between the CPU and GPU nodes on Cirrus.

%%%%%%%%%%%%%%%%%%%%%%%%%%%%%%%%%%%%%%%%%%%%%%
%%%%%%%%%%%%%%%%%%%%%%%%%%%%%%%%%%%%%%%%%%%%%%
%%%%%%%%%%%%%%%%%%%%%%%%%%%%%%%%%%%%%%%%%%%%%%

\section{Validation on real ASKAP observations} \label{sec:askap}

In this section, we revisit the ASKAP measurements of the field SB9442-35 previously imaged by \citet{wilber2023scalableI, wilber2023scalableII} with uSARA and AIRI. To evaluate the performance of proposed hyperspectral algorithms, Hyper-uSARA and HyperAIRI, we support our analysis with the reconstructions obtained by their monochromatic counterparts, as well as the hyperspectral CLEAN variant in WSClean.

\subsection{Data details for the field SB9442-35}

\begin{figure*}[t!]
    \centering
    \begin{tabular}{c@{\hspace{0.5\tabcolsep}}c}
         \includegraphics[width=0.48\textwidth]{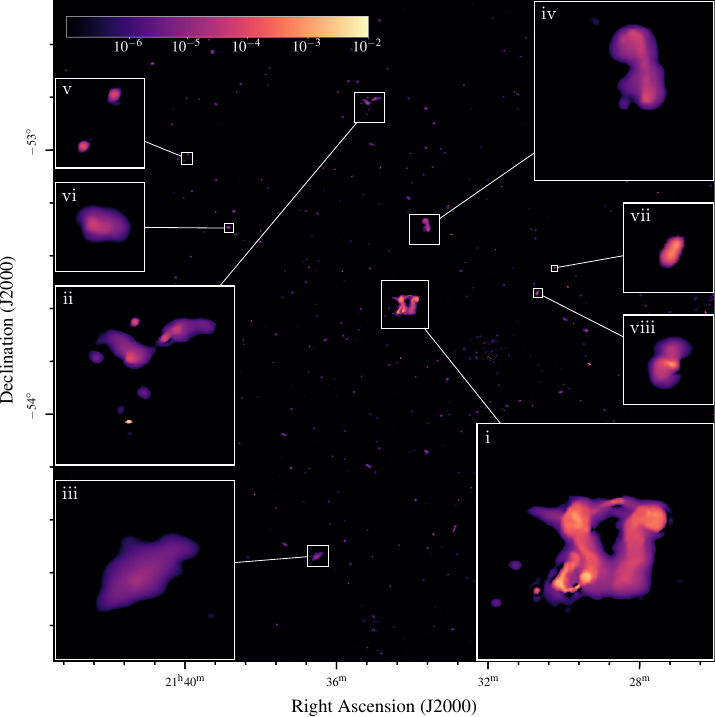}
         &  
        \includegraphics[width=0.48\textwidth]{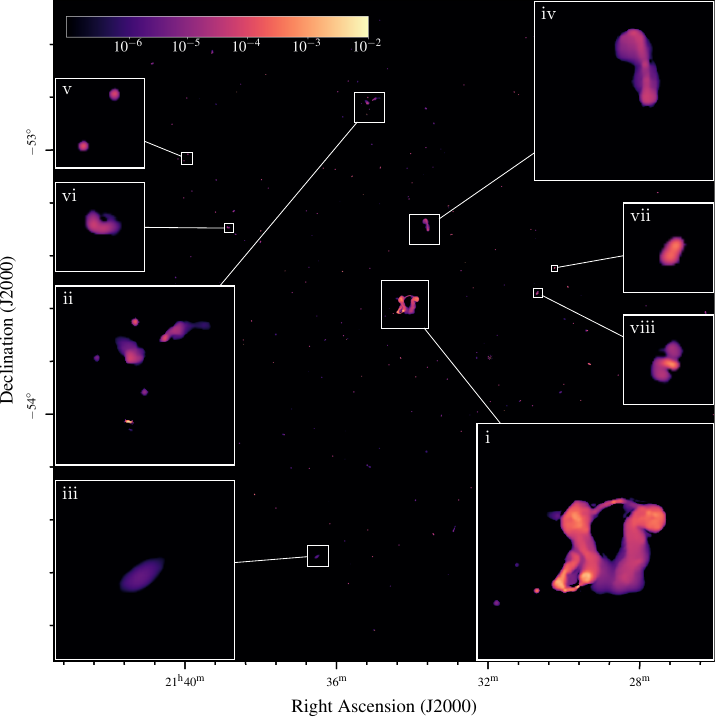}\\[-3pt]
         (a) AIRI, channel 1 image at 817 MHz &
         (b) AIRI, channel 8 image at 1069 MHz\\[4pt]
        \includegraphics[width=0.48\textwidth]{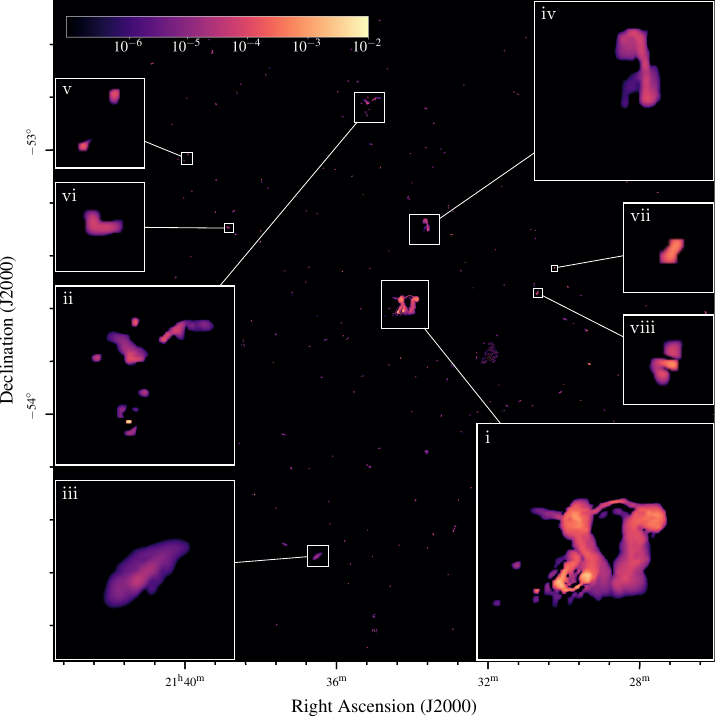}
         &  
        \includegraphics[width=0.48\textwidth]{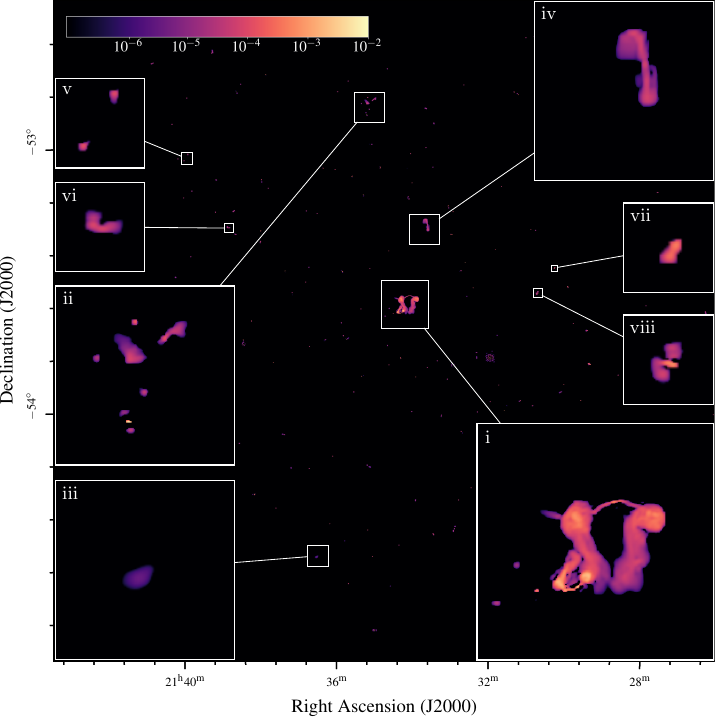}\\[-3pt]
         (c) uSARA, channel 1 image at 817 MHz &
         (d) uSARA, channel 8 image at 1069 MHz\\
    \end{tabular}
    \caption{ 
    Reconstructed images of the field SB9442-35 at the frequencies 817 MHz (a, c) and 1069 MHz (b, d), 
    obtained with AIRI (a, b), and uSARA (c, d) \citep{wilber2023scalableII, wilber2023scalableI}. Each panel is overlaid with zooms on selected regions: zoom (i) is centred on the ``dancing ghosts'', zooms (ii) and (iv) show regions with both extended and compact radio sources, zoom (iii) contains the star-forming galaxy NGC~7090, zooms (v)-(viii) focus on four compact sources.
    }
    \label{fig:askap_usara_airi}
\end{figure*}

\begin{figure*}[t]
    \centering
    \begin{tabular}{c@{\hspace{0.5\tabcolsep}}c}
        \includegraphics[width=0.48\textwidth]{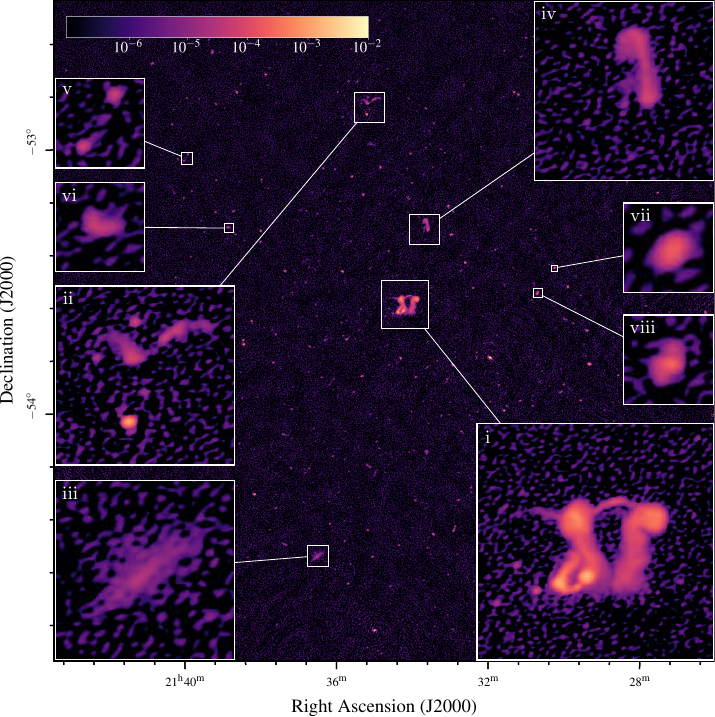}
         &  
        \includegraphics[width=0.48\textwidth]{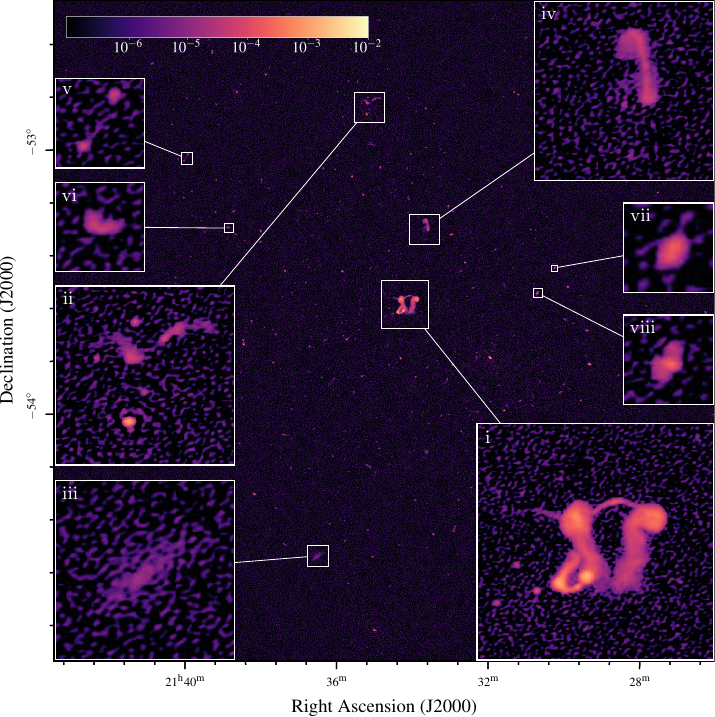}\\[-3pt]
         (a) WSClean, channel 1 image at 817 MHz &
         (b) WSClean, channel 8 image at 1069 MHz
    \end{tabular}
    \caption{Reconstructed images of the field SB9442-35 at the frequencies 817 MHz (a) and 1069 MHz (b), 
    obtained using  WSClean with joint-channel deconvolution. The zoom regions are identical to those in Figure~\ref{fig:askap_usara_airi}.
    }
    \label{fig:askap_wsclean}
\end{figure*}

The hyperspectral measurements used in this experiment are part of the ASKAP EMU Pilot Survey \citep{norris2021evolutionary}. In this survey, each Scheduling Block (SB) provides 36 different measurement sets, each corresponding to a different primary beam. For the real data experiment, we select the measurement set of Beam 35 of SB9442, centered at the radio source PKS~2130-538 of the galaxy cluster Abell~3785, known as the ``dancing ghosts''. 
The observation spans 10 hours covering the frequency band 800-1088 MHz with a frequency step size of 1 MHz, yielding 288 spectral channels. The selected hyperspectral measurements were calibrated with the ASKAPsoft \citep{hotan2021australian} and are accessible through the CSIRO ASKAP Science Data Archive \citep{chapman2015casda}.

\subsection{Imaging settings}

Following the imaging considerations of \citet{wilber2023scalableI, wilber2023scalableII}, the measurement set is divided into 8 spectral windows, combining 36 channels each.
The visibilities in each spectral window are concatenated together during imaging. The target image cube thus contains 8 spectral channels. Each channel has a pixel size of $4096 \times 4096$ and a cell size of 2.20 arcsec, achieving a final field of view of $2.57^\circ \times 2.57^\circ$, corresponding to a super-resolution factor of 2.22 at the highest-frequency channel and 2.91 at the lowest-frequency channel. The $w$-correction is enabled for all algorithms to account for $w$-effect in wide-field imaging. Briggs weighting with the robustness parameter fixed to $-0.25$ is considered. 

The choices of algorithm-specific parameters are detailed below. For WSClean, we enable the joint-channel deconvolution and consider 4 polynomial terms for spectral fitting.
The $w$-gridder is selected for gridding and $w$-correction.
The remaining parameters are set consistent with the considerations of \citet{wilber2023scalableI}, and the full WSClean command is shown in Appendix~\ref{sec:clean_cmd}.
For comparison, the WSClean image cube is normalized channel-wise by the flux of the corresponding restoring beam as per \eqref{eq:clean_beam_area}.
The regularization parameter $\lambda$ in Hyper-uSARA and the effective noise level for the denoisers in HyperAIRI are adjusted to 0.3 and 0.5 times the heuristic noise level \eqref{eq:heuristic} respectively. This adjustment improves the overall imaging quality while mitigating calibration artifacts. 
The MATLAB-based measurement operator encapsulates a hybrid $w$-projection $w$-stacking approach to correct for the $w$-effect \citep{dabbech2022first, dabbech2025distributed}. 

To evaluate the spectral index maps from the reconstructed image cubes, we follow the considerations of \citet{wilber2023scalableI, wilber2023scalableII}. For each image cube, a smoothing Gaussian kernel was applied, with a full width at half maximum (FWHM) of 20 arcsec for WSClean and 5 arcsec for the remaining algorithms.
Subsequently, the logarithm of the pixel intensities of these channels are fitted to a linear function with the slope to be the spectral index map, as shown in \eqref{eq:power_law_log}.

All experiments were run on the Cirrus HPC systems. For WSClean, 1 full CPU node is allocated. For Hyper-uSARA, 4 CPU nodes are assigned with 142 cores are used during reconstruction. Meanwhile, HyperAIRI ran on 4 GPU nodes, where 142 CPU cores and 8 GPUs are used.

\subsection{Reconstruction results}
\begin{figure*}[t]
    \centering
    \begin{tabular}{c@{\hspace{0.5\tabcolsep}}c}
         \includegraphics[width=0.48\textwidth]{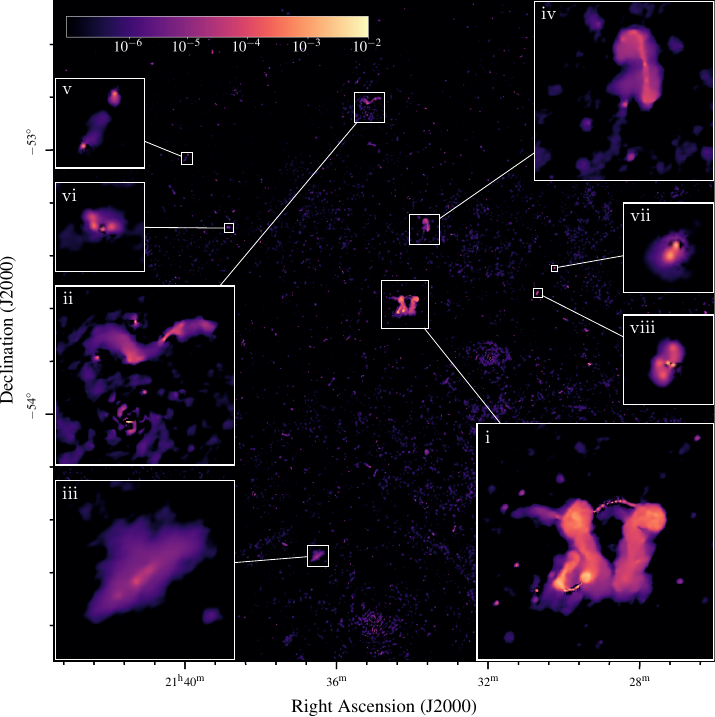}
         &  
        \includegraphics[width=0.48\textwidth]{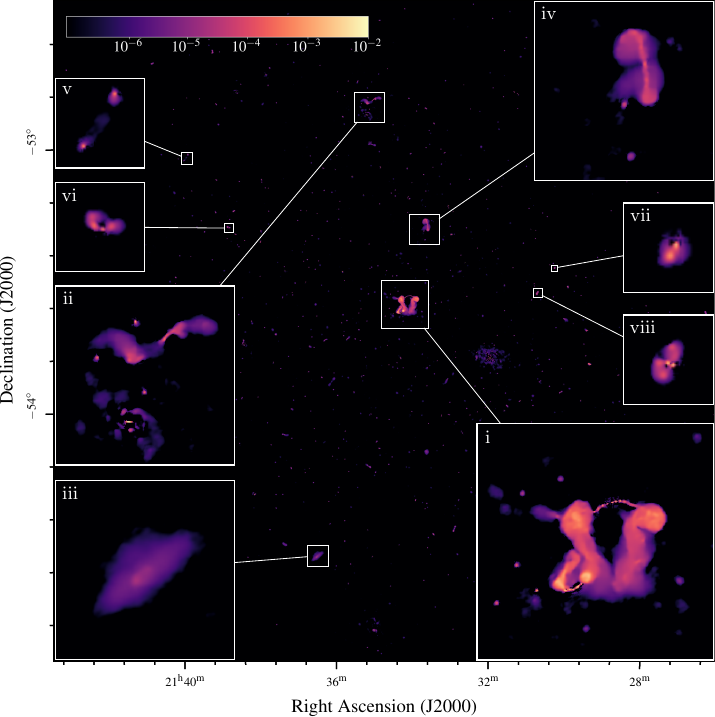}\\[-3pt]
        (a) HyperAIRI, channel 1 image at 817 MHz &
        (b) HyperAIRI, channel 8 image at 1069 MHz \\[4pt]
        \includegraphics[width=0.48\textwidth]{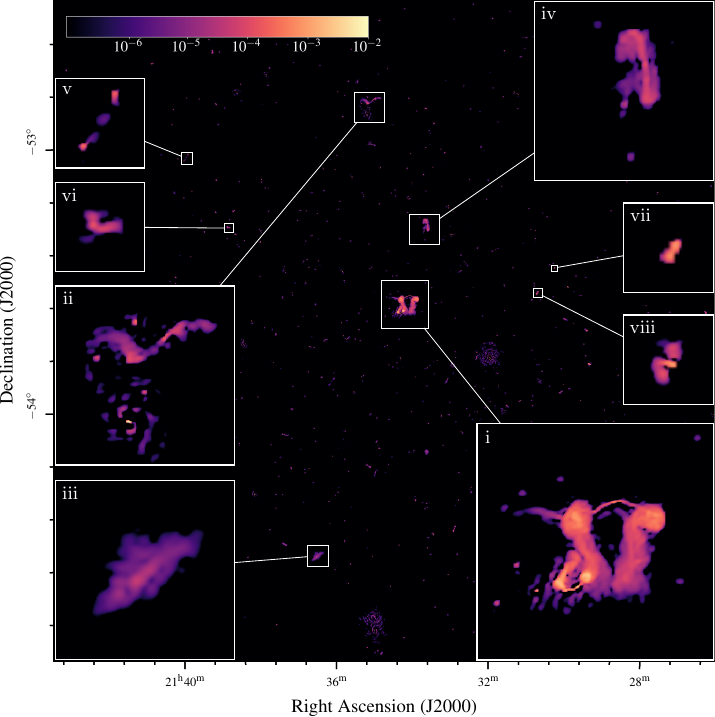}
        &  
        \includegraphics[width=0.48\textwidth]{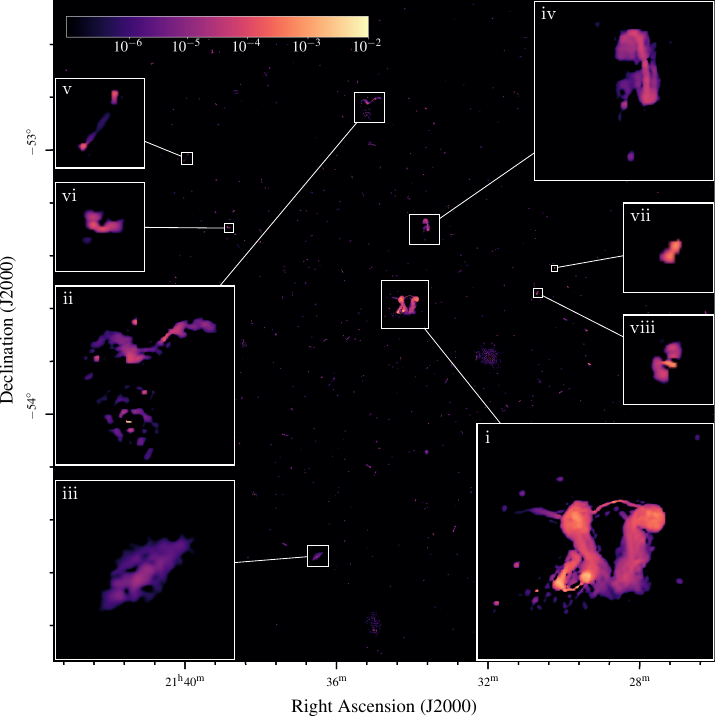}\\[-3pt]
        (c) Hyper-uSARA, channel 1 image at 817 MHz &
        (d) Hyper-uSARA, channel 8 image at 1069 MHz
    \end{tabular}
    \caption{ Reconstructed images of the field SB9442-35 at the frequencies 817 MHz (a, c) and 1069 MHz (b, d), 
    obtained with HyperAIRI (a, b) and Hyper-uSARA (c, d). 
    The zoom regions are identical to those in Figure~\ref{fig:askap_usara_airi}.}
    \label{fig:askap_hyperusara_hyperairi}
\end{figure*}

We examine the reconstructions provided by the different algorithms in Figure~\ref{fig:askap_usara_airi}-\ref{fig:askap_hyperusara_hyperairi}, at both lowest (panels (a) and (c)) and highest (panels (b) and (d)) frequencies. 
Figure~\ref{fig:askap_usara_airi} exhibits AIRI (panels (a) and (b)) and uSARA (panels (c) and (d)) images from \citet{wilber2023scalableII, wilber2023scalableI}.
Figure~\ref{fig:askap_wsclean} presents WSClean images. Figure~\ref{fig:askap_hyperusara_hyperairi} displays HyperAIRI (panels (a) and (b)) and Hyper-uSARA (panels (b) and (d)) images.
In each panel, we consider zooms on eight selected regions as follows: region (i) centered at the ``dancing ghosts''; regions (ii) and (iv) corresponding to areas containing both extended and compact radio sources; region (iii) centered on the star-forming galaxy NGC~7090; and regions (v)-(viii) centered on compact sources with varying morphologies.

Compared with their monochromatic counterparts, the proposed hyperspectral algorithms recover significantly more faint background sources, capture finer details of the radio emission, and exhibit improved inter-channel consistency. 
Among the hyperspectral approaches, HyperAIRI achieves distinguished source sharpness and recovers the largest number of compact sources, albeit with some noise-like residual patterns suggesting some extent of data over-fitting. 
On the other hand, Hyper-uSARA fails to capture some faint compact background sources due to its joint sparsity regularization, enforcing pixel values of sources, detected in some channels only, to be set to zero. WSClean reconstructions are smoother by design, with some faint sources appear to be buried in the noise.
While one could further optimize the noise-level scaling factors on a per-channel basis as in \citet{wilber2023scalableI, wilber2023scalableII} or consider per-channel visibility calibration to better balance background residuals against source recovery, these processing strategies lie beyond the scope of this experiment.

A detailed examination of the zoom regions highlights substantial differences in the imaging precision of the different algorithms in terms of angular resolution, dynamic range and the ability to capture faint emission, and spectral correlation.
Firstly, the superior angular resolution achieved by HyperAIRI at both channels compared to the benchmark is exemplified by zoom regions (vi), (vii) and (viii), with Hyper-uSARA ranking second. 
Both algorithms reveal inner cores and jets of the selected radio galaxies, whereas their monochromatic counterparts and even WSClean provide a blurred depiction of the same sources. Notably, HyperAIRI resolves the inner core of the radio galaxy in region (vi), and distinctly separates the two compact sources at the center of region (viii). 
Secondly, the enhanced sensitivity and dynamic-range of the proposed hyperspectral algorithms is showcased in zoom region (v). Both HyperAIRI and Hyper-uSARA successfully recover the faint jets emanating from the two hotspots. The faint features are completely absent in the monochromatic reconstructions, and are not revealed clearly in WSClean's images.  
Thirdly, consistency of spectral behavior achieved by the proposed hyperspectral algorithms is exemplified by the V-shaped emission in region (ii), the star-forming galaxy NGC~7090 in region (iii), and the two-sided jet emission and nearby point sources in region (iv), with faint structures preserved across channels.
Examining region (iii) at the highest frequency, monochromatic algorithms fail to recover NGC~7090's fusiform morphology. For WSClean, the source is buried in the noise.
By contrast, both HyperAIRI and Hyper-uSARA reliably restore NGC~7090's elongated form, with the former providing a clearer depiction of the diffuse emission.

\begin{figure*}
    \setlength{\fboxrule}{0.1pt}  % Thickness of the frame line
    \setlength{\fboxsep}{0pt}     % Gap between image and frame
    \centering
    \begin{tabular}{r @{\hspace{0.3\tabcolsep}} c @{\hspace{0.2\tabcolsep}} c @{\hspace{0.2\tabcolsep}} c @{\hspace{0.2\tabcolsep}} c @{\hspace{0.2\tabcolsep}} c @{\hspace{0.\tabcolsep}} c}
         & uSARA & AIRI & WSClean & Hyper-uSARA & HyperAIRI & \\
         \rotatebox[origin=c]{90}{Channel 1}&
         \raisebox{-0.5\height}{\fbox{\includegraphics[width=0.182\linewidth]{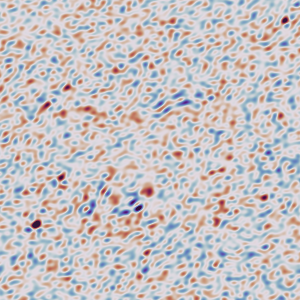}}}
         &
         \raisebox{-0.5\height}{\fbox{\includegraphics[width=0.182\linewidth]{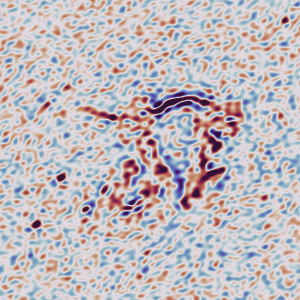}}}
         &
         \raisebox{-0.5\height}{\fbox{\includegraphics[width=0.182\linewidth]{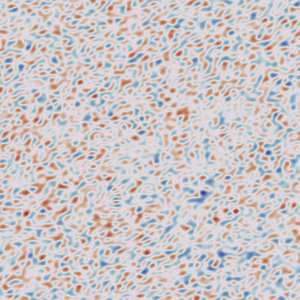}}}
         &
         \raisebox{-0.5\height}{\fbox{\includegraphics[width=0.182\linewidth]{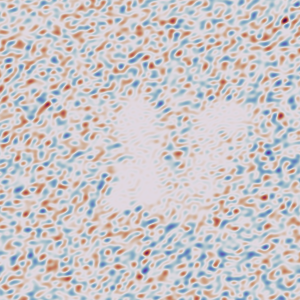}}}
         &
         \raisebox{-0.5\height}{\fbox{\includegraphics[width=0.182\linewidth]{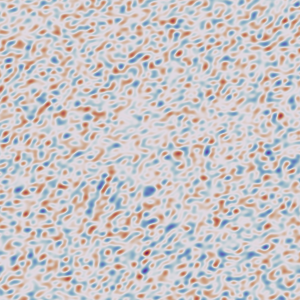}}} 
         &
         \raisebox{-0.5\height}{{\includegraphics[height=0.17\linewidth]{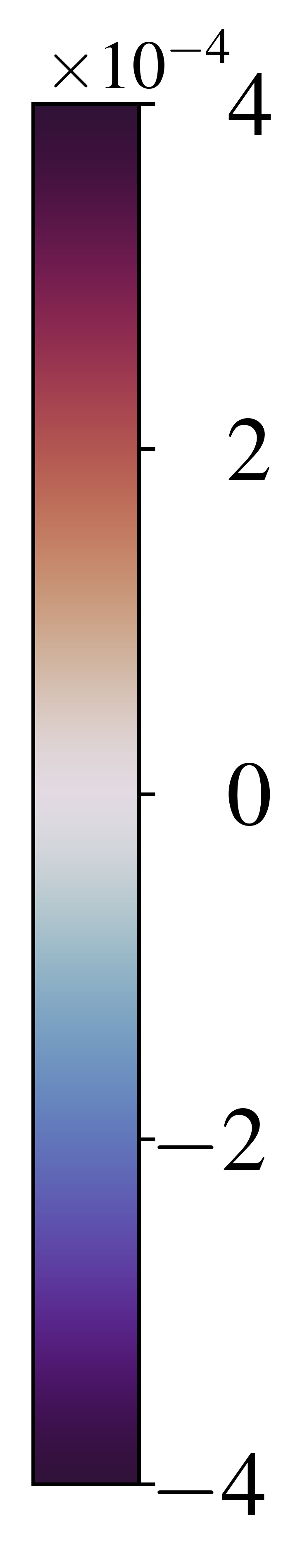}}}
         \\[1.57cm]
         \rotatebox[origin=c]{90}{Channel \added{8}}&
         \raisebox{-0.5\height}{\fbox{\includegraphics[width=0.182\linewidth]{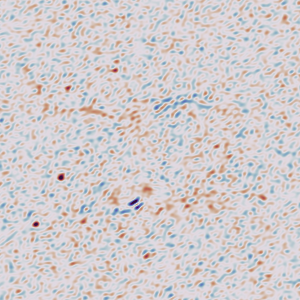}}}
         &
         \raisebox{-0.5\height}{\fbox{\includegraphics[width=0.182\linewidth]{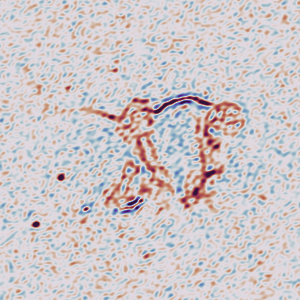}}}
         &
         \raisebox{-0.5\height}{\fbox{\includegraphics[width=0.182\linewidth]{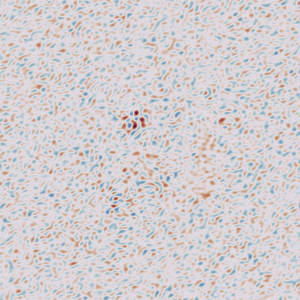}}}
         &
         \raisebox{-0.5\height}{\fbox{\includegraphics[width=0.182\linewidth]{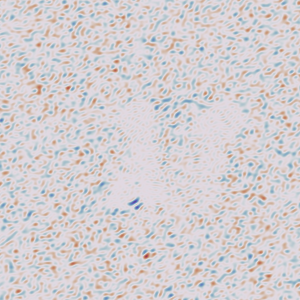}}}
         &
         \raisebox{-0.5\height}{\fbox{\includegraphics[width=0.182\linewidth]{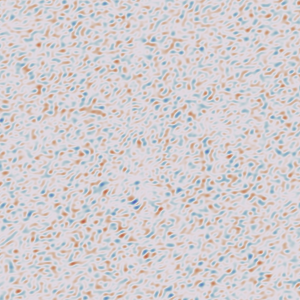}}} 
         &
         \raisebox{-0.5\height}{{\includegraphics[height=0.17\linewidth]{fig_askap_residual_cb_ver.png}}}
    \end{tabular}
    \caption{Residual dirty images zoomed at the ``dancing ghosts'', for channel 1 (top row) and 8 (bottom row) from the reconstructions produced by the algorithms shown in Figure~\ref{fig:askap_usara_airi}-\ref{fig:askap_hyperusara_hyperairi}}
    \label{fig:askap_residual}
\end{figure*}

\begin{figure*}
    \setlength{\fboxrule}{0.1pt}  % Thickness of the frame line
    \setlength{\fboxsep}{0pt}     % Gap between image and frame
    \centering
    \begin{tabular}{r @{\hspace{0.3\tabcolsep}} c @{\hspace{0.2\tabcolsep}} c @{\hspace{0.2\tabcolsep}} c @{\hspace{0.2\tabcolsep}} c @{\hspace{0.2\tabcolsep}} c @{\hspace{0.\tabcolsep}} c}
         \rotatebox[origin=c]{90}{\phantom{C}}&
         \fbox{\includegraphics[width=0.182\linewidth]{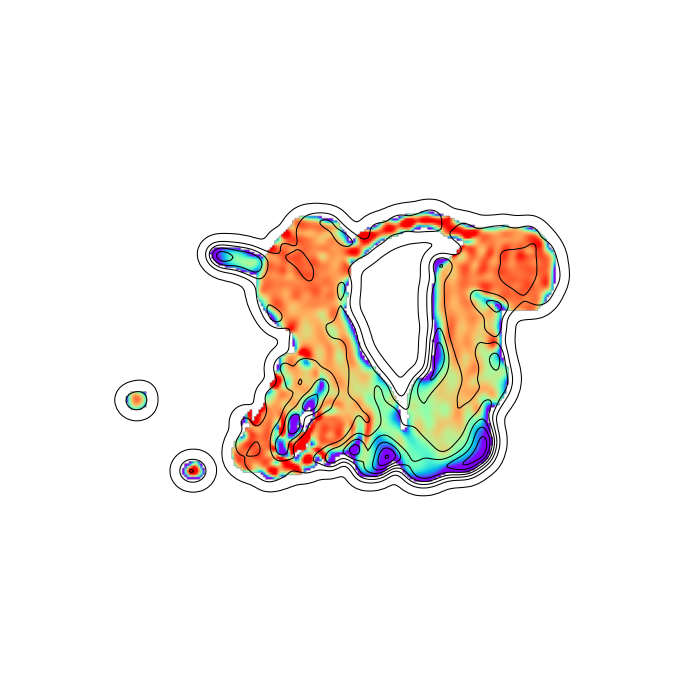}}
         &
         \fbox{\includegraphics[width=0.182\linewidth]{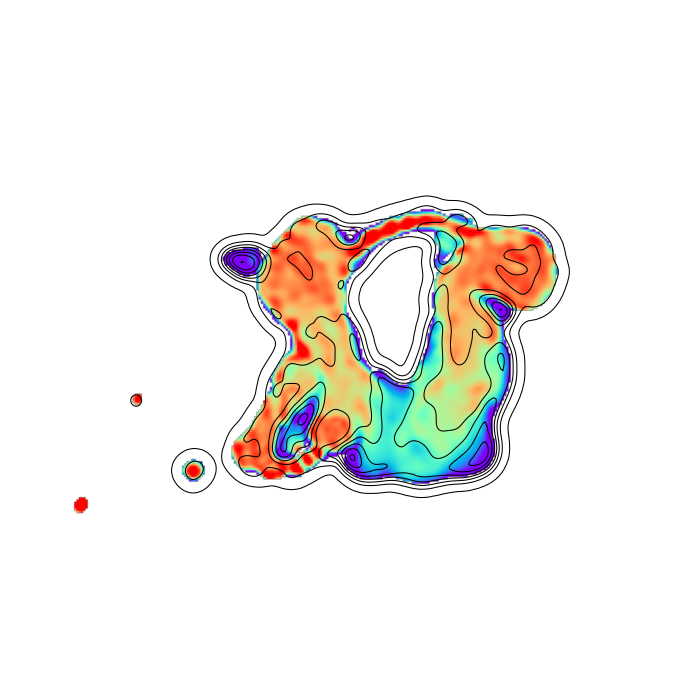}}
         &
         \fbox{\includegraphics[width=0.182\linewidth]{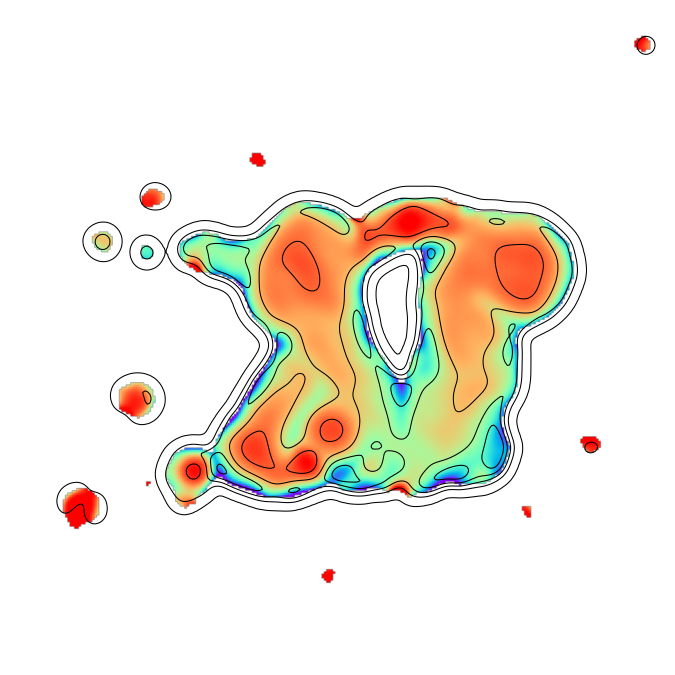}}
         &
         \fbox{\includegraphics[width=0.182\linewidth]{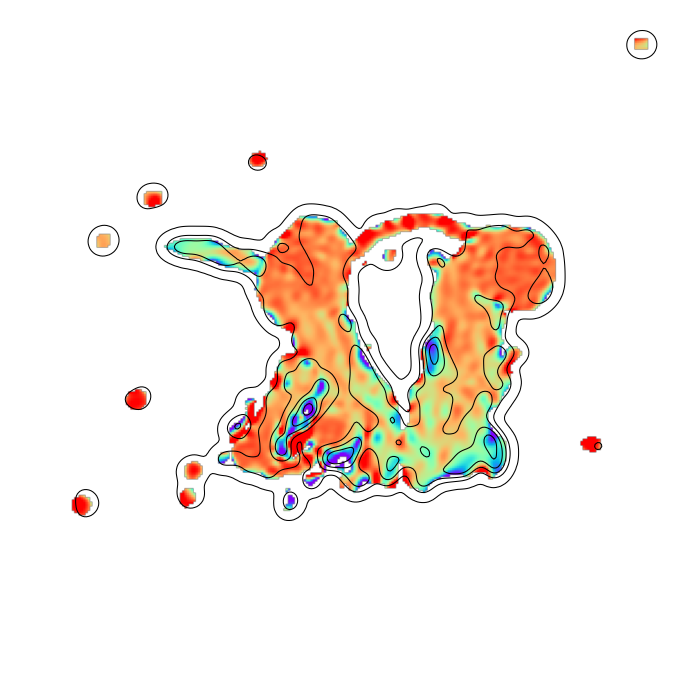}}
         &
         \fbox{\includegraphics[width=0.182\linewidth]{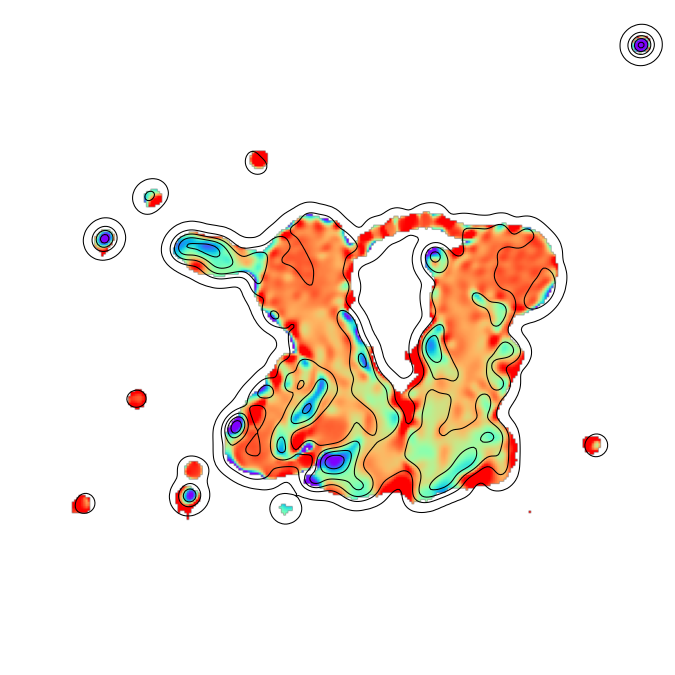}} 
         &
         \raisebox{2mm}{\includegraphics[height=0.16\linewidth]{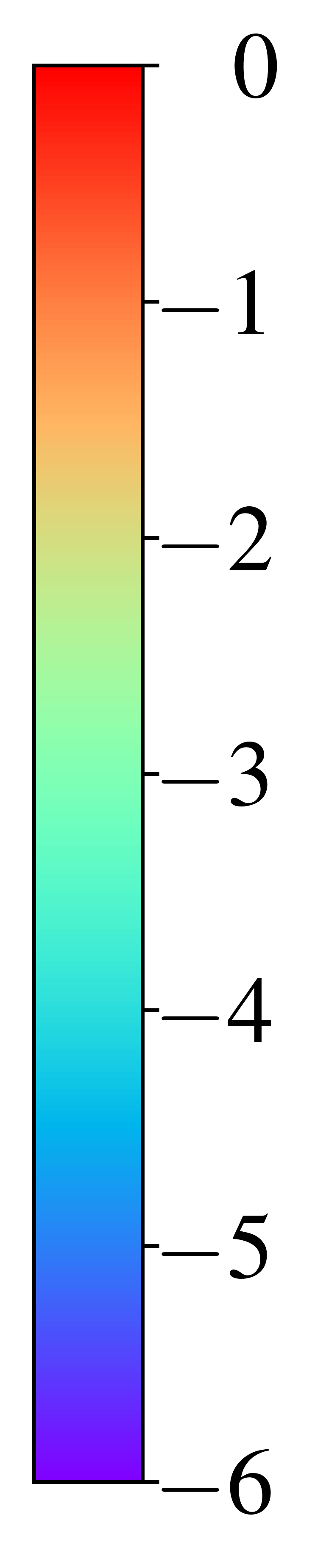}}
         \\
         & (a) uSARA & (b) AIRI & (c) WSClean & (d) Hyper-uSARA & (e) HyperAIRI &
    \end{tabular}
    \caption{The spectral index maps of the ``dancing ghosts'' evaluated from the reconstructions of different algorithms. Panels (a) and (b) are extracted from the reconstructions of \protect\cite{wilber2023scalableI, wilber2023scalableII}. The maps are computed after convolving channel images with a Gaussian beam of 20 arcsec for WSClean and 5 arcsec for the other reconstructions.}
    \label{fig:askap_spec_idx}
\end{figure*}

In what follows, we provide a deeper analysis of the recovery of the ``dancing ghosts'' in zoom region (i), including data fidelity and the source's spectral index map. The source is the most prominent in this field, notable for its complex morphology. It comprises the emission of two active galactic nuclei (AGN): the first (WISEA~J213406.70$-$533418.7) at the center of the north bridge, and the second (WISEA~J213417.69$-$533811.1) in the southeast filament \citep{norris2021evolutionary, velovic2023meerkat}. 
All hyperspectral algorithms provide consistent reconstructions across channels. In particular, two jets from the two lobes of the first AGN respectively pointing towards the northeast exhibit similar shapes in both the first and last channels.
Hyper-uSARA offers sharper edges and lower background noise, albeit with minor wavelets artifacts around the jets and lobes of the second AGN. HyperAIRI tends to deliver slightly higher resolution and recover more of the neighboring compact point-like sources than Hyper-uSARA. However, the PnP algorithm seems to depict the north bridge with a chain of dots, which is likely a reconstruction artifact. 
This could be indicative of training dataset bias, with half of the images simulated from optical astronomical images dominated by point-like sources. For insight into the achieved data fidelity, residual dirty images associated with zoom region (i) are shown in Figure~\ref{fig:askap_residual}. The hyperspectral algorithms deliver noise-like residuals, enabling improved data fidelity over monochromatic algorithms. 
Unlike HyperAIRI and WSClean, the signature of the ``dancing ghosts'' is somewhat visible in the residual dirty image of Hyper-uSARA, suggesting a minor data over-fitting.
The source's spectral index maps are shown in Figure~\ref{fig:askap_spec_idx}.
Hyperspectral algorithms yield smaller and more physically plausible spectral index values, while uSARA and AIRI show moderately higher values likely due to spectral inconsistency across channels. This is exemplified by the filaments associated with the two AGN and in their associated lobes.

Finally, the computational performance of the hyperspectral algorithms varies considerably due to the algorithmic and implementation differences. WSClean completes the reconstruction in 2.07 hours, making it the fastest. HyperAIRI requires 12.16 hours and runs 2500 iterations in FB, while Hyper-uSARA is the slowest, taking 24.48 hours to finish 1764 iterations in FB. Notably, the denoising (backward) step in Hyper-uSARA averages 35.13 seconds per iteration, compared to only 5.35 seconds for HyperAIRI. This difference is primarily due to the sub-iterative nature of the SARA-$\ell_{2,1}$ proximity operator and additional overhead from global gathering and scattering of wavelet coefficients for each channel during its denoising steps.

%%%%%%%%%%%%%%%%%%%%%%%%%%%%%%%%%%%%%%%%%%%%%%
%%%%%%%%%%%%%%%%%%%%%%%%%%%%%%%%%%%%%%%%%%%%%%
%%%%%%%%%%%%%%%%%%%%%%%%%%%%%%%%%%%%%%%%%%%%%%

\section{Conclusions} \label{sec:conclusion}

In this paper, we have introduced HyperAIRI, a novel PnP algorithm for extreme-scale hyperspectral RI imaging.
HyperAIRI builds on the monochromatic FB-PnP algorithm AIRI, adapting its DNN denoisers to process hyperspectral RI image cubes, thereby exploiting their inherent spectral correlations. 
We also formally introduce the optimization-based counterpart algorithm, Hyper-uSARA, which adopts the same FB iterative structure. Moreover, Hyper-uSARA can be interpreted as a simplified variant of HyperSARA \citep{abdulaziz2019wideband, thouvenin2023parallelI}, benefiting from an unconstrained formulation of the minimization task, thus eliminating the need for exact prior knowledge of the RI noise level.
 
HyperAIRI denoisers take as input adjacent spectral channels to capture local correlations, while simultaneously injecting a power-law spectral model that facilitates the propagation of global spectral information across channels.
This design reduces the need for global communications, thereby improving computational efficiency during both training and image reconstruction, while maintaining precision imaging.
Furthermore, HyperAIRI denoisers adapt easily to any image cube size through its flexible spatial-spectral faceting. Inheriting from AIRI its training strategy, which is fully decoupled from the measurement process, HyperAIRI denoisers can be applied under varying observational setup without the need for retraining. HyperAIRI also benefits from convergence guarantees owing to its non-expansive denoisers.  
The resulting precision, flexibility and scalability mark a significant improvement over the state-of-the-art methods, so far struggling to achieve these requirements simultaneously. 
 
These strengths of HyperAIRI are reflected in our experimental results both in simulation and on real observations. On simulated observations, it delivers substantial improvements in quantitative metrics, visual quality, data consistency, and spectral coherence compared with the monochromatic algorithms, AIRI and uSARA, as well as CLEAN with joint-channel deconvolution in WSClean. 
Results from real ASKAP observations further confirm the applicability of HyperAIRI in large-scale measurements. Compared with other methods, HyperAIRI is able to capture faint emissions, improve resolution, and enhance spectral correlation while maintaining high data fidelity.
 
Despite its promise for precision hyperspectral imaging, HyperAIRI remains computationally expensive due to the highly iterative nature of the underpinning FB, in which frequent calls of the measurement operator dominate the overall runtime. Inspired by CLEAN-based algorithms, exploring a major-minor cycle algorithmic structure with PSF approximations of the measurement operators inside minor cycles is a potential solution to reduce the computational cost. Additionally, building on the initial attempts in small-scale cases \citep{aghabiglou2025towards}, developing a fully GPU-based implementation tailored for extreme-scale problems can significantly bridge the computational gap with WSClean. 

Beyond scalability, further gains in reconstruction quality could be achieved. 
In particular, exploring advanced denoiser architectures that incorporate self- and cross-attention mechanisms \citep{vaswani2017attention}, to exploit non-local features across the spatial and spectral domains, could enhance HyperAIRI's precision.
Endowing HyperAIRI with a calibration functionality is also crucial to ensure the high-precision imaging required for high-sensitivity observations from next-generation telescopes.

%%%%%%%%%%%%%%%%%%%%%%%%%%%%%%%%%%%%%%%%%%%%%%
%%%%%%%%%%%%%%%%%%%%%%%%%%%%%%%%%%%%%%%%%%%%%%
%%%%%%%%%%%%%%%%%%%%%%%%%%%%%%%%%%%%%%%%%%%%%%

\section*{Data Availability}
HyperAIRI DNNs are available on the \href{https://doi.org/10.17861/d4412c73-ce64-431e-9728-0209a6b0ceb1}{Heriot-Watt University research portal}.
The images used to generate training datasets are sourced as follows. Optical astronomy images are gathered from NOIRLab/NSF/AURA/H.Schweiker/WIYN/T.A.Rector (University of Alaska Anchorage). Medical images are obtained from the NYU fastMRI Initiative database \citep{zbontar2018fastmri, knoll2020fastmri}.
The RI images used to build the simulated test dataset come from NRAO Archives, LOFAR HBA Virgo cluster survey \citep{edler2023victoria}, and LoTSS-DR2 survey \citep{shimwell2022vizier}.
The ASKAP data underlying this article (calibrated visibilities and mosaic images of Scheduling Blocks) are made publicly available for viewing and downloading at \href{https://data.csiro.au/collections/#domain/casdaObservation/search/}{CSIRO ASKAP Science Data Archive} \citep[CASDA,][]{chapman2015casda}, and can be accessed with the unique Project Identifiers AS101.
The uSARA and AIRI reconstructions of ASKAP used for comparison can be found through the DOI \href{https://doi.org/10.17861/5a60f25b-d43b-4807-ba02-a69bc460db03}{10.17861/5a60f25b-d43b-4807-ba02-a69bc460db03}.

\begin{acknowledgments}
This research was supported by UK Research and Innovation through the EPSRC grant EP/T028270/1, and 
the STFC grants ST/W000970/1 and APP31234. 
The research used Cirrus, a UK National Tier-2 HPC Service at EPCC funded by the University of Edinburgh and EPSRC (EP/P020267/1). ASKAP, from which the data under scrutiny originate, is part of the Australia Telescope National Facility managed by CSIRO. This project used public archival data from the Dark Energy Survey (DES).
\end{acknowledgments}

% \begin{contribution}

% All authors contributed equally to the Terra Mater collaboration.

%% But authors are expected to provide more specific details, e.g. 
%%
%%SC was responsible for writing and submitting the manuscript.
%%WWM came up with the initial research concept and edited the manuscript.
%%OTS obtained the funding and edited the manuscript.
%%EBF provided the formal analysis and validation. He also edited the manuscript.
%%GEH Supervised the undergraduates, wrote the software and administers the project github and Zenodo repositories.
%%
%% Authors can use the Contributor Role Taxonomy (CRediT) at
%% https://credit.niso.org
%% for ideas on how write a good statement tailored to their needs.

% \end{contribution}

%% To help institutions obtain information on the effectiveness of their 
%% telescopes the AAS Journals has created a group of keywords for telescope 
%% facilities.
%
%% Following the acknowledgments section, use the following syntax and the
%% \facility{} or \facilities{} macros to list the keywords of facilities used 
%% in the research for the paper.  Each keyword is check against the master 
%% list during copy editing.  Individual instruments can be provided in 
%% parentheses, after the keyword, but they are not verified.
\facilities{VLA, ASKAP.}

%% Similar to \facility{}, there is the optional \software command to allow 
%% authors a place to specify which programs were used during the creation of 
%% the manuscript. Authors should list each code and include either a
%% citation or url to the code inside ()s when available.
\software{
    % MATLAB \citep{MATLAB2023b},
    PyTorch \citep{paszke2019pytorch},
    NUFFT \citep{fessler2003nonuniform},
    % EIRA \citep{dabbech2025distributed},
    WSClean \citep{offringa2017optimized},
    MeqTrees \citep{noordam2010meqtrees},
    CASA \citep{bean2022casa}.
}

%% Appendix material should be preceded with a single \appendix command.
%% There should be a \section command for each appendix. Mark appendix
%% subsections with the same markup you use in the main body of the paper.
%%
%% Each Appendix (indicated with \section) will be lettered A, B, C, etc.
%% The equation counter will reset when it encounters the \appendix
%% command and will number appendix equations (A1), (A2), etc. The
%% Figure and Table counter will not reset.

\appendix

\section{Notation list}
\begin{table}[h]
    \centering
    \small
    \caption{Essential mathematical notations}
    \begin{tabular}{ll}
        \toprule 
        $(\cdot)_l$ 
        & 
        \begin{minipage}[t]{0.6\columnwidth}
            Choose $l$-th column from the input matrix 
        \end{minipage}
        \\
        $\|\cdot\|_1$, $\|\cdot\|_2$ 
        & 
        \begin{minipage}[t]{0.6\columnwidth}
            $\ell_1$ and $\ell_2$ norm 
        \end{minipage}
        \\
        $\|\cdot\|_\mathrm{F}$ & 
        \begin{minipage}[t]{0.6\columnwidth}
            Frobenius norm
        \end{minipage}
        \\
        $\|\cdot\|_\mathrm{S}$ 
        & 
        \begin{minipage}[t]{0.6\columnwidth}
            Spectral norm 
        \end{minipage}
        \\ 
        $\iota_\mathcal{C}(\cdot)$ 
        & 
        \begin{minipage}[t]{0.6\columnwidth}
            Indicator function of set $\mathcal{C}$ 
        \end{minipage}
        \\
        $\mathrm{prox}_r(\cdot)$
        & 
        \begin{minipage}[t]{0.6\columnwidth}
            Proximity operator for function $r(\cdot)$ 
        \end{minipage}
        \\
        $\mathrm{D}(\cdot)$ 
        & 
        \begin{minipage}[t]{0.6\columnwidth}
            Denoiser
        \end{minipage}
        \\
        $\Phi(\cdot), \Phi^\dagger(\cdot)$ 
        & 
        \begin{minipage}[t]{0.6\columnwidth}
            Forward and adjoint measurement operator for hyperspectral image cube 
        \end{minipage}
        \\
        \midrule
        $M$ & 
        \begin{minipage}[t]{0.6\columnwidth}
            Number of visibilities in each channel 
        \end{minipage}
        \\
        $N$ & 
        \begin{minipage}[t]{0.6\columnwidth}
            Number of pixels in the image of each channel 
        \end{minipage}
        \\
        $L$ 
        & 
        \begin{minipage}[t]{0.6\columnwidth}
            Total number of channels 
        \end{minipage}
        \\
        $b$ 
        & 
        \begin{minipage}[t]{0.6\columnwidth}
            Number of wavelet bases for the redundant wavelet dictionary 
        \end{minipage}
        \\
        \midrule
        $\bm{\Phi}_l$, $\bm{\Phi}_l^\dagger$ 
        & 
        \begin{minipage}[t]{0.6\columnwidth}
            Forward and adjoint measurement operator at channel $l$
        \end{minipage}
        \\
        $\mathbf{X}=\left ( \left [ \bm{x}_l \right ]_{1 \leq l \leq L}  \right )$ 
        & 
        \begin{minipage}[t]{0.6\columnwidth}
            Hyperspectral image cube
        \end{minipage}
        \\
        $\overline{\mathbf{X}}$ 
        & 
        \begin{minipage}[t]{0.6\columnwidth}
            Ground truth image cube
        \end{minipage}
        \\
        $\mathbf{X}^{\mathrm{dirty}}$ 
        & 
        \begin{minipage}[t]{0.6\columnwidth}
            Dirty image cube
        \end{minipage}
        \\
        $\mathbf{X}^{\mathrm{res}}$ 
        & 
        \begin{minipage}[t]{0.6\columnwidth}
            Residual dirty image cube
        \end{minipage}
        \\
        $\mathbf{Y}=\left ( \left [ \bm{y}_l \right ]_{1 \leq l \leq L}  \right )$ 
        & 
        \begin{minipage}[t]{0.6\columnwidth}
            Hyperspectral measurements
        \end{minipage}
        \\
        $\mathbf{N}=\left ( \left [ \bm{n}_l \right ]_{1 \leq l \leq L}  \right )$ 
        & 
        \begin{minipage}[t]{0.6\columnwidth}
            Hyperspectral measurements noise
        \end{minipage}
        \\
        $\tau_l$ 
        & 
        \begin{minipage}[t]{0.6\columnwidth}
            Standard deviation of $\bm{n}_l$
        \end{minipage}
        \\
        $\nu_l$ 
        & 
        \begin{minipage}[t]{0.6\columnwidth}
            Spectral frequency of channel $l$ 
        \end{minipage}
        \\
        $\breve{\bm{\nu}}_l = (\breve{{\nu}}_{l-1,l}, \breve{{\nu}}_{l+1,l})$ 
        & 
        \begin{minipage}[t]{0.6\columnwidth}
            The frequency ratios between $\nu_{l-1}$, $\nu_{l+1}$ and $\nu_{l}$ respectively 
        \end{minipage}
        \\
        $\bm{\alpha}$, $\bm{\beta}$ 
        &
        \begin{minipage}[t]{0.6\columnwidth} 
            Spectral index map and spectral curvature map
        \end{minipage}
        \\
        $\widetilde{\bm{\alpha}}$ 
        &
        \begin{minipage}[t]{0.6\columnwidth} 
            Estimated spectral index map
        \end{minipage}
        \\
        $\sigma_{\mathrm{heu},l}$ 
        & 
        \begin{minipage}[t]{0.6\columnwidth}
            Heuristic noise level in the image domain of channel $l$
        \end{minipage}
        \\
        $\rho_l$ 
        & 
        \begin{minipage}[t]{0.6\columnwidth}
            Maximum pixel intensity of image in channel $l$
        \end{minipage}
        \\
        $\bm{\Psi}^\dagger$, $\bm{\Psi}$ 
        & 
        \begin{minipage}[t]{0.6\columnwidth}
            Redundant wavelet dictionary and its adjoint transform 
        \end{minipage}
        \\
        $\lambda$ 
        &
        \begin{minipage}[t]{0.6\columnwidth} 
            The regularization parameter in the objective function
        \end{minipage}
        \\
        $\gamma$ 
        & 
        \begin{minipage}[t]{0.6\columnwidth}
            The step size for the gradient descent
        \end{minipage}
        \\
        $\epsilon$ 
        & 
        \begin{minipage}[t]{0.6\columnwidth}
            The estimated noise floor level in the wavelet coefficients
        \end{minipage}
        \\
        \bottomrule
    \end{tabular}
    \label{tab:notations}
\end{table}
A list of mathematical notations used in this paper is listed in Table.~\ref{tab:notations}.

\section{WSClean Imaging Command}\label{sec:clean_cmd}

We include the exact WSClean commands used to generate the reference reconstructions in our experiments. The parameters were chosen to ensure consistent spectral modelling and high imaging fidelity for both the simulated and ASKAP datasets.
In simulated experiment, we used the following command
\begin{verbatim}
wsclean -join-channels -no-mf-weighting \
  -fit-beam -fit-spectral-pol 5 \
  -channel-range 0 36 -channels-out 36 \
  -multiscale -scale ${PIXELSIZE}asec \ 
  -niter 6000000 -nmiter 30 -weight \
  briggs 0.0 -mgain 0.8 -size 512 512 \
  -auto-threshold 0.4 -minuvw-m 0.001 \
  -auto-mask 1 -mem 95 ${MSPATH}.
\end{verbatim}
The pixel size in the command is calculated from the desired super-resolution factor.
For the ASKAP data, the WSClean command is
\begin{verbatim}
wsclean -join-channels -gridder wgridder \
  -size 4096 4096 -scale 2.2asec \
  -minuvw-m 60 -no-mf-weighting \
  -fit-spectral-pol 4 -channel-range 0 288 \
  -channels-out 8 -mem 95 \
  -weight briggs -0.25 \
  -weighting-rank-filter 3 \
  -reorder -clean-border 1 -multiscale \
  -no-min-grid-resolution \
  -auto-threshold 1.0 -auto-mask 2.5 \
  -mgain 0.8 -fit-beam -pol i \
  -niter 2000000 -name SB9442-35 ${MSPATH}.
\end{verbatim}

%% For this sample we use BibTeX plus aasjournalv7.bst to generate the
%% the bibliography. The sample7.bib file was populated from ADS. To
%% get the citations to show in the compiled file do the following:
%%
%% pdflatex sample7.tex
%% bibtext sample7
%% pdflatex sample7.tex
%% pdflatex sample7.tex

\bibliographystyle{aasjournalv7}
\bibliography{hyperairi}

%% This command is needed to show the entire author+affiliation list when
%% the collaboration and author truncation commands are used.  It has to
%% go at the end of the manuscript.
%\allauthors

%% Include this line if you are using the \added, \replaced, \deleted
%% commands to see a summary list of all changes at the end of the article.
%\listofchanges

\end{document}